\documentclass[12pt]{iopart}
\usepackage{iopams}
\usepackage{pstricks,pst-node,pst-text,pst-3d,pst-plot}
\usepackage{bm}          
\def\today{Feb 22, 2010 \\ Revised: Apr 26, 2010} 
\begin{document}

\newcommand{\be}{\begin{equation}}
\newcommand{\ee}{\end{equation}} 
\newcommand{\<}{\langle}
\renewcommand{\>}{\rangle}
\newcommand{\widebar}{\overline}
\def\reff#1{(\protect\ref{#1})}
\def\spose#1{\hbox to 0pt{#1\hss}}
\def\ltapprox{\mathrel{\spose{\lower 3pt\hbox{$\mathchar"218$}}
 \raise 2.0pt\hbox{$\mathchar"13C$}}}
\def\gtapprox{\mathrel{\spose{\lower 3pt\hbox{$\mathchar"218$}}
 \raise 2.0pt\hbox{$\mathchar"13E$}}}
\def\textprime{${}^\prime$} 
\def\proof{\par\medskip\noindent{\sc Proof.\ }}
\def\qed{\hbox{\hskip 6pt\vrule width6pt height7pt depth1pt \hskip1pt}\bigskip}
\def\proofof#1{\bigskip\noindent{\sc Proof of #1.\ }}
\def\half{ {1 \over 2} }
\def\third{ {1 \over 3} }
\def\twothird{ {2 \over 3} }
\def\smfrac#1#2{\textstyle{\frac{#1}{#2}}}
\def\smhalf{ \smfrac{1}{2} }
\newcommand{\real}{\mathop{\rm Re}\nolimits}
\renewcommand{\Re}{\mathop{\rm Re}\nolimits}
\newcommand{\imag}{\mathop{\rm Im}\nolimits} 
\renewcommand{\Im}{\mathop{\rm Im}\nolimits}
\newcommand{\sgn}{\mathop{\rm sgn}\nolimits}
\newcommand{\diag}{\mathop{\rm diag}\nolimits}
\newcommand{\Gal}{\mathop{\rm Gal}\nolimits}
\newcommand{\mycup}{\mathop{\cup}}
\newcommand{\Arg}{\mathop{\rm Arg}\nolimits}
\def\hboxscript#1{ {\hbox{\scriptsize\em #1}} }
\def\zhat{ {\widehat{Z}} } 
\def\phat{ {\widehat{P}} }
\def\qtilde{ {\widetilde{q}} }
\newcommand{\mod}{\mathop{\rm mod}\nolimits}
\renewcommand{\emptyset}{\varnothing}

\def\scra{\mathcal{A}}
\def\scrb{\mathcal{B}}
\def\scrc{\mathcal{C}}
\def\scrd{\mathcal{D}}
\def\scrf{\mathcal{F}}
\def\scrg{\mathcal{G}}
\def\scrl{\mathcal{L}}
\def\scro{\mathcal{O}}
\def\scrp{\mathcal{P}}
\def\scrq{\mathcal{Q}}
\def\scrr{\mathcal{R}}
\def\scrs{\mathcal{S}}
\def\scrt{\mathcal{T}}
\def\scrv{\mathcal{V}}
\def\scrz{\mathcal{Z}}

\def\q{{\sf q}}

\def\Z{{\mathbb Z}}
\def\R{{\mathbb R}}
\def\C{{\mathbb C}}
\def\Q{{\mathbb Q}}
\def\N{{\mathbb N}}

\def\T{{\mathsf T}}
\def\H{{\mathsf H}}
\def\V{{\mathsf V}}
\def\D{{\mathsf D}}
\def\J{{\mathsf J}}
\def\P{{\mathsf P}}
\def\QQ{{\mathsf Q}}
\def\RR{{\mathsf R}}

\def\bsigma{{\boldsymbol{\sigma}}}
\def\bone{{\mathbf 1}}
\def\vv{{\bf v}}
\def\uu{{\bf u}}
\def\w{{\bf w}}

\newtheorem{theorem}{Theorem}[section]
\newtheorem{definition}[theorem]{Definition}
\newtheorem{proposition}[theorem]{Proposition}
\newtheorem{lemma}[theorem]{Lemma}
\newtheorem{corollary}[theorem]{Corollary}
\newtheorem{conjecture}[theorem]{Conjecture}


\newenvironment{sarray}{
          \textfont0=\scriptfont0
          \scriptfont0=\scriptscriptfont0
          \textfont1=\scriptfont1
          \scriptfont1=\scriptscriptfont1
          \textfont2=\scriptfont2
          \scriptfont2=\scriptscriptfont2
          \textfont3=\scriptfont3
          \scriptfont3=\scriptscriptfont3
        \renewcommand{\arraystretch}{0.7}
        \begin{array}{l}}{\end{array}}

\newenvironment{scarray}{
          \textfont0=\scriptfont0
          \scriptfont0=\scriptscriptfont0
          \textfont1=\scriptfont1
          \scriptfont1=\scriptscriptfont1
          \textfont2=\scriptfont2
          \scriptfont2=\scriptscriptfont2
          \textfont3=\scriptfont3
          \scriptfont3=\scriptscriptfont3
        \renewcommand{\arraystretch}{0.7}
        \begin{array}{c}}{\end{array}}

\newcommand\Kc{\kappa}  
\newcommand\proofofcase[2]{\bigskip\noindent{\sc Case #1: #2.\ }}
\newcommand\algcr{\mathop{algcr}}

\bibliographystyle{plain}

\date{\today}

\title[On the non-ergodicity of the WSK algorithm]%
{On the non-ergodicity of the Swendsen--Wang-\\Koteck\'y algorithm on
    the kagom\'e lattice} 

\author{Bojan Mohar$^1$\footnote{On leave from Department of Mathematics, 
              IMFM \& FMF, University of Ljubljana, Ljubljana, Slovenia.}
              and Jes\'us Salas$^2$} 
\address{$^1$ Department of Mathematics, Simon Fraser University, 
              Burnaby, B.C.~~V5A 1S6, Canada.}
\ead{mohar@sfu.ca}

\medskip

\address{$^2$ Instituto Gregorio Mill\'an and 
              Grupo de Modelizaci\'on, Simulaci\'on Num\'erica y Matem\'atica
              Industrial, Universidad Carlos III de Madrid, 
              Avda.\  de la Universidad, 30, 28911 Legan\'es, Spain.} 
\ead{jsalas@math.uc3m.es}

\thispagestyle{empty}   

\begin{abstract}
We study the properties of the Wang--Swendsen--Koteck\'y cluster Monte Carlo
algorithm for simulating the 3-state kagom\'e-lattice Potts antiferromagnet 
at zero temperature. We prove that this algorithm is not ergodic for symmetric 
subsets of the kagom\'e lattice with fully periodic boundary conditions: 
given an initial configuration, not all configurations are accessible 
via Monte Carlo steps. The same conclusion holds for single-site dynamics.  
\end{abstract}

\pacs{02.10.Ox 02.40.Re 02.50.Ga 05.50.+q 64.50.De}
\maketitle

%
%
\section{Introduction}

The $q$-state Potts model \cite{Potts_52,Wu_82,Wu_84} is certainly one of
the simplest and most studied models in Statistical Mechanics.
However, despite many efforts over more than 50 years, its {\em exact}
solution (even in two dimensions) is still unknown. The ferromagnetic
regime is the best understood case: there are exact (albeit not always
rigorously proved) results for the location of the critical temperature,
the order of the transition, etc.
The antiferromagnetic regime is less understood, partly because
universality is not expected to hold in general (in contrast with the
ferromagnetic regime case); in particular, critical behaviour may depend
on the lattice structure of the model.
One interesting feature of this antiferromagnetic
regime is that it exhibits non-zero ground-state entropy (without frustration)
for large enough values of $q$ on a given lattice. This provides an 
exception to the third law of thermodynamics \cite{Lieb,Chow}. 
In addition, a zero-temperature phase transition may occur for certain values
of $q$ and certain lattices: e.g., the models with $q=2,4$ on the triangular
lattice, and $q=3$ on the square and kagom\'e lattices, 
cf.\ \cite{Salas_Sokal_97} and references therein. 

In addition to its intrinsic theoretical interest, the antiferromagnetic 
3--state Potts model on the kagom\'e lattice also plays an important role in 
condensed--matter physics. Several experimental systems are proposed to 
be modelled by antiferromagnetic $n$-component $O(n)$ spin models  
on the kagom\'e lattice 
\cite[and references therein]{Chalker,Huse,Ritchey,Henley}. 
For both the XY ($n=2$) and Heisenberg ($n=3$) models, there are
theoretical arguments showing that, in the zero-temperature limit, 
their ground states can be described 
by the 3--state antiferromagnetic Potts model. Furthermore, Huse and 
Rutenberg \cite{Huse} showed that exactly at zero temperature, the later 
model has a SOS (or height) representation, and it is critical. 

The standard $q$-state Potts model can be defined on any finite undirected
graph $G = (V,E)$ with vertex set $V$ and edge set $E$. On each vertex
$i\in V$ of the graph $G$, we place a spin $\sigma(i)\in \{1,2,\ldots,q\}$,
where $q\ge 2$ in an integer. The spins interact via a Hamiltonian
\begin{equation}
H(\{\sigma\}) \;=\; -J \sum\limits_{e=ij\in E} \delta_{\sigma(i),\sigma(j)} \,,
\end{equation}
where the sum is over all edges $e \in E$, $J\in\R$ is the coupling constant,
and $\delta_{a,b}$ is the Kronecker delta.
The {\em Boltzmann weight}\/ of a configuration is then $e^{-\beta H}$,
where $\beta \ge 0$ is the inverse temperature.
The {\em partition function}\/ is the sum, taken over all configurations,
of their Boltzmann weights:
\begin{equation}
   Z_G^{\rm Potts}(q, \beta J)  \;=\;  
   \sum_{ \sigma \colon\, V \to \{ 1,2,\ldots,q \} } \; 
      e^{- \beta H(\{\sigma\}) } \,.
 \label{def.ZPotts}
\end{equation}
A coupling $J$ is called {\em ferromagnetic}\/ if $J \ge 0$,
as it is then favored for adjacent spins to take the same value; and
{\em antiferromagnetic}\/ if $-\infty \le J \le 0$,
as it is then favored for adjacent spins to take different values.
The zero-temperature ($\beta \to +\infty$) limit of the antiferromagnetic
($J < 0$) Potts model has an interpretation as a colouring problem: the limit
$\lim_{\beta\to+\infty} Z_G^{\rm Potts}(q,-\beta |J|)=P_G(q)$ is the
{\em chromatic polynomial}, which gives the number of proper
$q$-colourings of $G$. A {\em proper $q$-colouring}\/ of $G$ is a map
$\sigma \colon\, V \to \{ 1,2,\ldots,q \}$ such that
$\sigma(i) \neq \sigma(j)$ for all pairs of adjacent vertices $ij\in E$.
In other words, a proper $q$-colouring of a graph $G$ is a colouring of the 
vertices of $G$ such that any pair of nearest-neighbour vertices are not 
coloured alike. 

For many Statistical Mechanics systems for which an exact solution is
not known, (Markov chain) Monte Carlo simulations \cite{Bremaud,Landau}
have become a very valuable tool to extract physical information.
One popular Monte Carlo algorithm for the {\em antiferromagnetic}\/
$q$-state Potts model is the Wang--Swendsen--Koteck\'y (WSK) {\em non-local}
cluster dynamics \cite{WSK_89,WSK_90}. Even though at any positive 
temperature the WSK algorithm satisfies {\em all}\/  the necessary 
conditions in order to work, {\em exactly}\/  at zero temperature, one 
condition (i.e., ergodicity) may not hold, and therefore, the algorithm 
can no longer be used! A Monte Carlo algorithm is {\em ergodic}\/  (or 
irreducible) if it can eventually get from each state (or configuration)
to every other state. 
While this condition is easy to check for the WSK algorithm at any 
positive temperature, it becomes a highly non-trivial question at 
zero temperature for non-bipartite graphs.  

It is interesting to note that at zero temperature, the basic moves of
the WSK dynamics correspond to the so-called {\em Kempe changes},
introduced by Kempe in his unsuccessful proof of the four-colour theorem
\cite[Section~7.3.1]{Gibbons} (see also \cite{Wagon} anc check 
\cite{Mohar_05} for additional references). This 
zero-temperature algorithm (disguised under the name of `path--flipping' 
algorithm) has already been used by several authors \cite{Huse,Ritchey}. 
In particular,
Huse and Rutenberg \cite{Huse} noted (see their footnote~13) that for fully 
periodic boundary conditions this algorithm is not ergodic; but we are 
not aware of any (rigorous) proof of this claim in the literature. 

It is also worth noticing that for $q$-state Potts antiferromagnets without
frustration (e.g., $q \ge 3$ for the kagom\'e lattice), single--spin flips 
are a (proper) subset of the set of Kempe  moves. Therefore, 
the non-ergodicity of the later dynamics implies the non-ergodicity of  
single--flip algorithms (which include the well--known Metropolis 
algorithm \cite{Landau}). For positive temperature, the WSK algorithm (on 
any graph) always include single-site moves as a special case.

Although the Potts model can be defined on any graph $G$, in
Statistical Mechanics one is mainly interested in `large' regular graphs
with fully periodic boundary conditions (i.e., embedded on a torus). 
Boundary conditions of this type are usually chosen to minimize 
finite-size-scaling effects \cite{Privman}. Therefore, we will focus on 
the commonest set-up in actual Monte Carlo simulations: finite 
symmetric subsets of the kagom\'e lattice wrapped on a torus. 

The ergodicity of the WSK algorithm for the zero-temperature 
$q$-state Potts antiferromagnet on the kagom\'e lattice embedded 
on a torus is only an open question for $q=3,4$. 
For $q=2$ (the Ising model) it is trivially non-ergodic, as each
WSK move is equivalent to a global spin flip. 
It is interesting to remark that there is an analytic solution for the 
Ising model on the kagom\'e lattice \cite{Kano_53}; this solution 
shows that there is no phase transition in the whole antiferromagnetic 
regime, including zero temperature, where the system displays frustration.
On the contrary, for $q\ge 5$ the algorithm is ergodic (see 
Section~\ref{sec.setup} for more details). 
Among the two unknown cases, $q=3$ is the most interesting one, because 
the system is expected to be critical at zero temperature~\cite{Huse}.

The main result of this paper is to provide a {\em proof}\/ of the 
non-ergodicity of the zero--temperature WSK algorithm for the
3--state Potts antiferromagnet on symmetric subsets
of the kagom\'e lattice with fully periodic boundary conditions. 
We find that the ground-state configuration space (i.e., the set of all
proper 3-colourings of the given kagom\'e graph) can be split into 
{\em at least} two `ergodicity classes' (or Kempe equivalence classes),
such that one class is unreachable using Kempe moves from the other one, 
and vice versa.    
This also means that single--flip dynamics is also non-ergodic for 
such systems. In this case, each ground-state configuration constitutes
an ergodicity class. 
Therefore, these zero-temperature Monte Carlo algorithms
simply do not work, and new algorithms satisfying all the required 
properties should be sought in order to simulate such systems.  
Furthermore, no reasonable algorithm is known at present to our knowledge,
which is ergodic at zero temperature. It is an interesting open problem
to find one. 

Our basic strategy in this paper goes as follows: We start with the 
observation that the kagom\'e lattice is the medial of the triangular 
lattice. \footnote{
  See Section~\ref{sec.new2b} for a precise definition of the medial 
  graph $M(G)$ of a graph $G$.
} 
In particular, for the reasons explained above, 
we are interested in the kagom\'e graphs $T'(3L,3L)$ which
are the medials of the regular triangulations $T(3L,3L)$ of the torus (roughly 
speaking, these triangulations are subsets of the triangular lattice 
of linear size $(3L)\times(3L)$ with fully periodic boundary conditions). 
We then show that any proper 3--colouring $\phi$ of the kagom\'e lattice 
$T'(3L,3L)$ can be viewed as a particular proper 4--colouring $f$ 
of the `doubled' 
triangulation $T(6L,6L)$, and that any WSK transition made on $\phi$ 
corresponds to a sequence of WSK moves performed on $f$ in $T(6L,6L)$. 
We call the colourings $f$ of $T(6L,6L)$ that are obtained from 3--colourings 
of the kagom\'e lattice {\em special colourings\/} of $T(6L,6L)$.
This correspondence enables us to use the results of Ref.~\cite{Mohar_Salas} 
about the non-ergodicity of the zero-temperature WSK algorithm for the 
4--state Potts antiferromagnet on the triangulations $T(3L,3L)$ with $L\ge 2$.

Proper 4--colourings of a triangulation embedded on a torus are rather
special, as they can be regarded as maps from a sphere to the torus 
(using the tetrahedral representation of the spin). This
basic observation allowed us to borrow concepts from algebraic topology;
in particular, the degree $\deg(f)$ of a proper 4--colouring $f$. 
This approach was pioneered by Fisk \cite{Fisk_73a,Fisk_77a,Fisk_77b}, who 
also showed that $\deg(f)$ on any 3--colourable triangulation of the torus
is always a multiple of 6 (the triangulations $T(3L,3L)$ are indeed 
3--colourable). 
We then showed that $\deg(f)\pmod{12}$ is an invariant under a Kempe move. 
Therefore, if we are able to find two 4--colourings $f$ and $g$
with degrees $\deg(f)\pmod{12}=0$ and $\deg(g)\pmod{12}=6$, then there  
are {\em al least}\/ two Kempe equivalence classes, and therefore the
zero--temperature WSK algorithm is not ergodic. For all triangulations
$T(3L,3L)$ with $L\ge 2$ we were able to find such two 4--colourings  
\cite{Mohar_Salas}.

In this paper, we apply these results to the subset of special 
4--colourings of $T(6L,6L)$: for any $L\ge 1$, we find that there are two
special 4--colourings $f$ and $g$ with degrees congruent with
$0$ and $6$ modulo $12$, respectively. Therefore, we cannot get $f$ from $g$
(or vice versa) using Kempe moves, even in the larger configuration space of
all proper 4--colourings of $T(6L,6L)$. The same is true if we restrict 
ourselves to the smaller set of {\em special}\/ proper 4--colourings of 
$T(6L,6L)$, which corresponds to the set of proper 3--colourings of 
$T'(3L,3L)$. Therefore, the non-ergodicity of the Kempe dynamics for 
these kagom\'e graphs follows. 
 
As explained above, our approach, based on algebraic topology, can only   
be applied to proper 4--colourings of the triangulations $T(3L,3L)$, or 
to proper 3--colourings of the kagom\'e graphs $T'(3L,3L)$ (by exploiting
that $T'$ is the medial of $T$). Unfortunately, it cannot be extended
to study the ergodicity of the WSK algorithm for the 4--state 
kagom\'e--lattice antiferromagnet. It is curious that our methods work for
the two models that have a height representation and are critical at 
zero temperature \cite{Huse}. It would be interesting to translate our 
findings into the height language \cite{Henley}. 
This may lead to a improved (and hopefully) ergodic algorithm.  

Finally, one might consider simulating the 3--state Potts antiferromagnet 
using the WSK algorithm at a small but {\em positive}\/ temperature. In 
this case, the algorithm is indeed ergodic and satisfies all the required
properties to work fine. However, the only way we can reach from one
ergodicity class to another is through a non-zero-energy configuration
(or non--proper 3--colouring). But these configurations are exponentially
suppressed in this limit: we have to pay a penalty of $e^{-\beta|J|}$ for
each pair of neighbouring vertices coloured alike. Therefore, it is unlikely
for small enough temperatures that the system visit more than one class.
Furthermore, it would be very interesting to consider, in addition to the
standard observables, new observables specifically designed to `feel' 
the nonergodicity of the algorithm, and to study numerically how their 
autocorrelation times behave as we approach to zero temperature. 

The paper is organized as follows: In Section~\ref{sec.setup} we
introduce our basic definitions, and review what is
known in the literature about the problem of the ergodicity of the
Kempe dynamics. 
In Section~\ref{sec.new2b} we consider edge-colourings and relate them first
to three-colourings of the kagom\'e graph and then to special colourings of
triangulations $T(6L,6M)$.
In Section~\ref{sec.main} we prove our main result about 
the non-ergodicity for symmetric kagom\'e graphs $T'(3L,3L)$. 

%
%
\section{Basic Setup} \label{sec.setup}

Let $G = (V,E)$ be a finite undirected graph with vertex set $V$ and edge set
$E$. Then for each graph $G$ there exists
a polynomial $P_G$ with integer coefficients such that, for each $q \in \Z_+$,
the number of proper $q$-colourings of $G$ is precisely $P_G(q)$.
This polynomial $P_G$ is called the {\em chromatic polynomial}\/ of $G$.
The set of all proper $q$-colourings of $G$ will be denoted as
$\mathcal{C}_q = \mathcal{C}_q(G)$ (thus, $|\mathcal{C}_q(G)|=P_G(q)$).

It is far from obvious that $Z_G^{\rm Potts}(q, \beta J)$
[cf. \reff{def.ZPotts}], which is defined separately for each positive
integer $q$, is in fact the restriction to $q \in \Z_+$ of
a {\em polynomial}\/ in $q$. But this is in fact the case, and indeed we have:

\begin{theorem}[Fortuin--Kasteleyn \protect\cite{Kasteleyn_69,Fortuin_72}
   representation of the Potts model] \label{thm.FK}
For every integer $q \ge 1$, we have
\begin{equation}
   Z_G^{\rm Potts}(q, v) \;=\;  
   \sum_{ A \subseteq E }  q^{k(A)} \, v^{|A|} \;,
 \label{eq.FK.identity}
\end{equation}
where $v=e^{\beta J}-1$, and $k(A)$ denotes the number of connected components
in the spanning subgraph $(V,A)$.
\end{theorem}

The foregoing considerations motivate defining the {\em Tutte polynomial}\/
of the graph $G$:
\begin{equation}
   Z_G(q, v)   \;=\;
   \sum_{A \subseteq E}  q^{k(A)} \,  v^{|A|} \;,
 \label{def.ZG}
\end{equation}
where $q$ and $v$ are commuting indeterminates. This polynomial is
equivalent to the standard Tutte polynomial $T_G(x,y)$ after a simple
change of variables. If we set $v=-1$, we obtain the
{\em chromatic polynomial} $P_G(q) = Z_G(q,-1)$. In particular, $q$ and
$v$ can be taken as complex variables. See Ref.~\cite{Sokal_bcc2005} for a
recent survey.

\begin{figure}[htb]
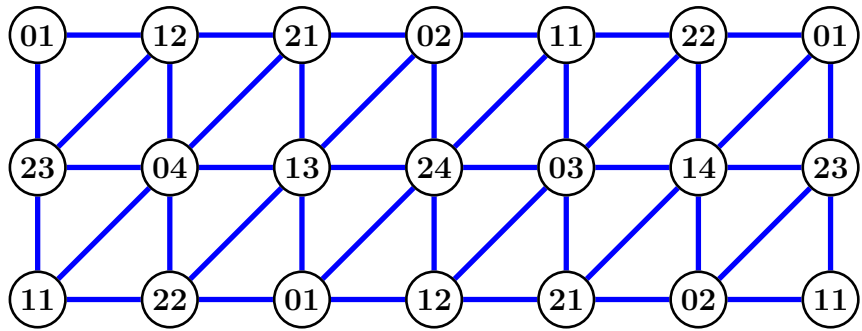

\centering
\psset{xunit=50pt}
\psset{yunit=50pt}
\psset{labelsep=10pt}
\pspicture(-0.5,-0.5)(6.5,2.5)
\multirput{0}(0,0)(0,1){3}{\psline[linewidth=2pt,linecolor=blue](0,0)(6,0)}
\multirput{0}(0,0)(1,0){7}{\psline[linewidth=2pt,linecolor=blue](0,0)(0,2)}
\multirput{0}(0,0)(1,0){5}{\psline[linewidth=2pt,linecolor=blue](0,0)(2,2)}
\psline[linewidth=2pt,linecolor=blue](0,1)(1,2)
\psline[linewidth=2pt,linecolor=blue](5,0)(6,1)
\multirput{0}(0,0)(0,1){3}{%
   \multirput{0}(0,0)(1,0){7}{%
      \pscircle*[linecolor=white]{11pt}
      \pscircle[linewidth=1pt,linecolor=black]{11pt}
   }
}
\rput{0}(0,0){$\bm{11}$}
\rput{0}(1,0){$\bm{22}$}
\rput{0}(2,0){$\bm{01}$}
\rput{0}(3,0){$\bm{12}$}
\rput{0}(4,0){$\bm{21}$}
\rput{0}(5,0){$\bm{02}$}
\rput{0}(6,0){$\bm{11}$}

\rput{0}(0,2){$\bm{01}$}
\rput{0}(1,2){$\bm{12}$}
\rput{0}(2,2){$\bm{21}$}
\rput{0}(3,2){$\bm{02}$}
\rput{0}(4,2){$\bm{11}$}
\rput{0}(5,2){$\bm{22}$}
\rput{0}(6,2){$\bm{01}$}

\rput{0}(0,1){$\bm{23}$}
\rput{0}(1,1){$\bm{04}$}
\rput{0}(2,1){$\bm{13}$}
\rput{0}(3,1){$\bm{24}$}
\rput{0}(4,1){$\bm{03}$}
\rput{0}(5,1){$\bm{14}$}
\rput{0}(6,1){$\bm{23}$}
\endpspicture
\caption{\label{figure_T_6_2_2}
The triangulation $T(6,2,2)=\Delta^2 \times \partial\Delta^3$ of the torus. 
Each vertex $x$ of $T(6,2,2)$ is labelled by two integers $ij$,
where  $i$ (resp.\ $j$) corresponds to the associated vertex in 
$\Delta^2$ (resp.\ $\partial\Delta^3$).
The vertices of $\Delta^2$ are labelled $\{0,1,2\}$, while the vertices of
$\partial\Delta^3$ are labelled $\{1,2,3,4\}$. The triangulation
$T(6,2,2)$ has $12$ vertices, and those in the figure with the same label
should be identified. 
}
\end{figure}

As explained in the Introduction, we will focus on kagom\'e lattices that
are related to certain regular triangulations embedded on the torus.
The class of regular triangulations of the torus with degree six is
characterized by the following theorem:

\begin{theorem}[Altschulter \protect\cite{Altschulter}] 
Let\/ $T$ be a triangulation of the torus such that all vertices have degree 
six. Then\/ $T$ is one of triangulations $T(r,s,t)$, which are obtained
from the $(r+1)\times (s+1)$ grid by adding diagonals in the squares of
the grid as shown in Figure~\ref{figure_T_6_2_2}, and then identifying
opposite sides to get a triangulation of the torus.
In $T(r,s,t)$ the top and bottom rows have $r$ edges, the left and right
sides $s$ edges. The left and right side are identified as usual; 
but the top and the bottom row are identified after (cyclically) shifting
the top row by $t$ edges to the right. 
\end{theorem}

%
%
%
%

In Figure~\ref{figure_T_6_2_2} we have displayed the triangulation $T(6,2,2)$
of the torus. We will represent these triangulations as
embedded on a rectangular
grid with three kinds of edges: horizontal, vertical, and diagonal.
The three-colourability of the triangulations $T(r,s,t)$ is given by the
following result \cite{Mohar_Salas}:

\begin{proposition}
The triangulation $T(r,s,t)$ is three-colourable if and only if
$r\equiv 0 \pmod{3}$ and $s-t\equiv 0 \pmod{3}$.
\end{proposition}

In Monte Carlo simulations, it is usual to consider toroidal boundary conditions
with no shifting, so $t=0$. Then, the three-colourability condition reduces
to the standard result $r,s\equiv 0\pmod{3}$. In general, we will consider
the following triangulations of the torus $T(3L,3M,0)=T(3L,3M)$ with
$L,M\geq 1$. The unique three-colouring $c_0$ of $T(3L,3M)$ can be described as:
\be
c_0(x,y) \;=\; \mod(x+y-2,3) + 1 \,,
               \quad 1\leq x \leq 3L\,, \quad 1\leq y \leq 3M \,,
\label{def_colouring_c0}
\ee
where we have explicitly used the above-described embedding of the
triangulation $T(3L,3M)$ in a square grid.

Finally, in most Monte Carlo simulations one usually considers tori of
aspect ratio one: i.e., $T(3L,3L)$. This is the class of triangulations we
are most interested in from the point of view of Statistical Mechanics.

%
%
\subsection{Kempe changes}

Given a graph $G=(V,E)$ and $q\in\N$, we can define the following dynamics on 
$\mathcal{C}_q$: Choose uniformly at random two distinct colours 
$a,b\in\{1,2,\ldots,q\}$, and let $G_{ab}$ be the induced subgraph of $G$ 
consisting of vertices $x\in V$ for which $\sigma(x)\in \{a,b\}$. Then, 
independently for each connected component of $G_{ab}$, with probability 
$1/2$ either interchange the colours $a$ and $b$ on it, or leave 
the component unchanged. 
This dynamics is the zero--temperature limit of the 
Wang--Swendsen--Koteck\'y (WSK) cluster dynamics \cite{WSK_89,WSK_90} for 
the antiferromagnetic $q$-state Potts model.
This zero--temperature Markov chain leaves invariant the uniform
measure over proper $q$-colourings; but its irreducibility cannot be 
taken for granted. 

The basic moves of the WSK dynamics correspond to Kempe changes
(or K-{\em changes}). In each K-change, we interchange the colours $a,b$ on
a given connected component (or K-{\em component\/}) of the induced subgraph 
$G_{ab}$.
 
Two $q$-colourings $c_1,c_2\in\mathcal{C}_q(G)$ related by a series of 
K-changes are {\em Kempe equivalent\/} (or K$_q$-{\em equivalent\/}). 
This (equivalence) relation is denoted as $c_1 \stackrel{q}{\sim} c_2$. 
The equivalence classes $\mathcal{C}_q(G)/\stackrel{q}{\sim}$ are called the
{\em Kempe classes\/} (or {\em K$_q$-classes\/}). The number of K$_q$-classes 
of $G$ is denoted by $\Kc(G,q)$. Then, if $\Kc(G,q)>1$, the 
zero-temperature WSK dynamics is not ergodic on $G$ for $q$ colours.

In this paper, we will consider two $q$-colourings related by a {\em global}
colour permutation to be the same. In other words, a $q$-colouring is
actually an equivalence class of standard $q$-colourings modulo global 
colour permutations. Thus, the number of (equivalence classes of) proper
$q$-colourings is given by $P_G(q)/q!$. This convention will simplify 
the notation in the sequel.

%
%
\subsection{The number of Kempe classes}

In this section we will briefly review what it is known in the literature 
about the number of Kempe equivalence classes for several families of graphs. 
The first result implies that WSK dynamics is ergodic on any bipartite 
graph:\footnote{All the cited authors have discovered this theorem 
independently.} 

\begin{proposition}[Burton \& Henley \protect\cite{Henley_97a}, %
 Ferreira \& Sokal \protect\cite{Sokal_99a}, %
 Mohar \protect\cite{Mohar_05}]
$\!$\label{prop.bipartite} \newline
Let\/ $G$ be a bipartite graph and $q\geq 2$ an integer. Then, $\Kc(G,q)=1$.
\end{proposition} 
It is worth noting that Lubin and Sokal \cite{Sokal_93} showed that 
the WSK dynamics with 3 colours is not ergodic on any square--lattice 
grid of size $3M\times 3N$ (with $M,N$ relatively prime) wrapped on a torus.
These graphs are indeed non-bipartite. 

The second type of results deals with graphs of bounded maximum degree 
$\Delta$, and shows that $\Kc(G,q)=1$ whenever $q$ is large enough:

\begin{proposition}[Jerrum \protect\cite{Jerrum_private} and %
 Mohar \protect\cite{Mohar_05}]
\label{prop.deltamax}
Let $\Delta$ be the maximum degree of a graph $G$ and let $q\geq \Delta+1$ be
an integer. Then $\Kc(G,q)=1$. If $G$ is connected and contains a vertex 
of degree $<\Delta$, then also $\Kc(G,\Delta)=1$.
\end{proposition}
This result implies that for any kagom\'e lattice $T'$ with $\Delta=4$, 
$\Kc(T',q)=1$ for any $q\geq \Delta+1=5$. Notice that the cases $q=2,3,4$ 
are not covered by the above proposition. 

Finally, if we consider planar graphs the situation is better 
understood. One of the authors proved that  

\begin{theorem}[Mohar \protect\cite{Mohar_05}, Theorem~4.4]
Let $G$ be a three-colourable planar graph. Then $\Kc(G,4)=1$.
\end{theorem}

\begin{corollary}[Mohar \cite{Mohar_05}, Corollary~4.5]
Let $G$ be a planar graph and $q > \chi(G)$. Then $\Kc(G,q)=1$.
\end{corollary}
These results imply that WSK for $q\ge 4$ is ergodic on any 
three-colourable planar graph. But we cannot use these results, as 
none of our graphs is planar. 

The main theorem for triangulations appears in Ref.~\cite{Fisk_77b} 
and involves the notion of the degree of a four-colouring, whose definition
is deferred to the next section.

\begin{theorem}[Fisk \protect\cite{Fisk_77b}] \label{theo_Fisk}
Suppose that\/ $T$ is a triangulation of the sphere, projective plane, 
or torus. If\/ $T$ has a three-colouring, then all four-colourings with 
degree divisible by $12$ are Kempe equivalent.
\end{theorem}

In a previous paper \cite{Mohar_Salas}, we proved a series of 
results that are of great importance in the present work.
The first theorem ensures the existence of a Kempe invariant for the class of
three-colourable triangulations of a closed orientable surface. 

\begin{theorem} \label{main.theo}
Let\/ $T$ be a three-colourable triangulation of a closed orientable surface.
If $f$ and $g$ are two four-colourings of\/ $T$ related by a Kempe change on a
region $R$, then
\be
\deg(g) \;\equiv \;\deg(f) \pmod{12} \,.
\label{main.eq}
\ee
\end{theorem}
Note that the class of three-colourable triangulations of a closed 
orientable surface contains and it is wider than the class $T(3L,3M)$ we 
are interested in. This theorem and Fisk's theorem \ref{theo_Fisk}, 
imply the following Corollary:  

\begin{corollary}
\label{main.corollary}
Let\/ $T$ be a three-colourable triangulation of the torus.
Then $\Kc(T,4)>1$ if and only if there exists a four-colouring $f$ with
$\deg(f)\equiv 6 \pmod{12}$.
\end{corollary}

For symmetric triangulations $T(3L,3L)$ we were able to prove the following
result: 
\begin{theorem} \label{theo.main}
For any triangulation $T(3L,3L)$ with $L\geq 2$ there exists a four-colouring
$f$ with $\deg(f)\equiv 6 \pmod{12}$. Hence, $\Kc(T(3L,3L),4)> 1$.
In other words, the WSK dynamics for four-colourings on $T(3L,3L)$
is non-ergodic.
\end{theorem}

For non-symmetric triangulations $T(3L,3M)$ our results can be summarized 
in the following theorem: 

\begin{theorem} \label{theo.asym}
For any triangulation $T(3L,3M)$ with any $L\geq 3$ and $M\geq L$,
there exists a four-colouring $f$ with $\deg(f)\equiv 6\pmod{12}$.
Consequently, the WSK dynamics for four-colourings of $T(3L,3M)$ is
non-ergodic.
\end{theorem}

For triangulations $T(6,3M)$ with $M\ge 2$, we could only prove the 
non-ergodicity of the WSK dynamics for $q=4$ when $M=2p$ with odd $p$, 
while this dynamics is ergodic at least for the triangulation $T(6,9)$.
Finally, we also proved that the WSK dynamics for $q=4$ is always
ergodic on any triangulation of the type $T(3,3M)$ with $M\ge 1$.

%
%
\section{Edge-colourings of triangulations of the torus}
\label{sec.new2b}

Four-colourings of triangulations of the 2-dimensional sphere are in
a bijective correspondence with three other kinds of colourings: edge-colourings, 
Heawood colourings, and local colourings. When treated on the torus (or on any
orientable surface of positive genus), these notions are no longer equivalent
to each other, but there is a nice hierarchy among them as shown by
Fisk \cite{Fisk_77b}. Under this hierarchy, 
every 4-colouring induces an edge-colouring, every
edge-colouring induces a Heawood colouring, every Heawood colouring induces
a local colouring, and all these correspondences are 1-1. 
However, none of these implications can be reversed.

\begin{figure}[ht]
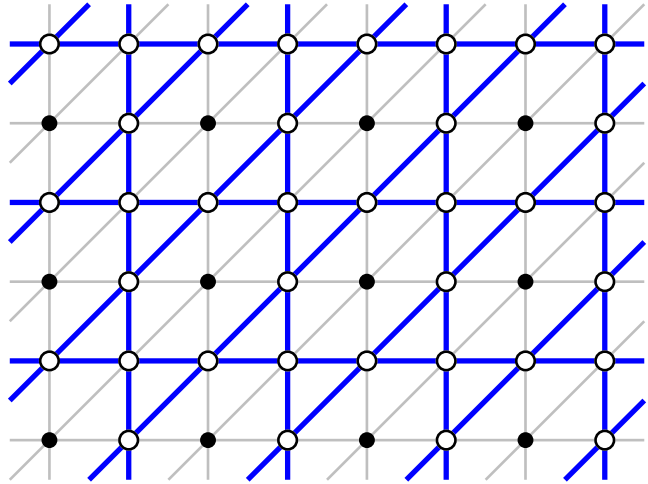

\centering
%
%
\psset{xunit=30pt}
\psset{yunit=30pt}
\pspicture(-1,-1)(8,6)
\multirput{0}(0,0)(2,0){4}{%
 \psline[linewidth=1pt,linecolor=lightgray](0,-0.5)(0,5.5)
}
\multirput{0}(0,0)(0,2){3}{%
 \psline[linewidth=1pt,linecolor=lightgray](-0.5,0)(7.5,0)
}
\psline[linewidth=1pt,linecolor=lightgray](-0.5, 3.5)(1.5,5.5)
\psline[linewidth=1pt,linecolor=lightgray](-0.5, 1.5)(3.5,5.5)
\psline[linewidth=1pt,linecolor=lightgray](-0.5,-0.5)(5.5,5.5)
\psline[linewidth=1pt,linecolor=lightgray]( 1.5,-0.5)(7.5,5.5)
\psline[linewidth=1pt,linecolor=lightgray]( 3.5,-0.5)(7.5,3.5)
\psline[linewidth=1pt,linecolor=lightgray]( 5.5,-0.5)(7.5,1.5)
\multirput{0}(0,1)(0,2){3}{%
  \psline[linewidth=2pt,linecolor=blue](-0.5,0)(7.5,0)
}
\multirput{0}(1,0)(2,0){4}{%
  \psline[linewidth=2pt,linecolor=blue](0,-0.5)(0,5.5)
}
\psline[linewidth=2pt,linecolor=blue](0.5,-0.5)(6.5,5.5)
\psline[linewidth=2pt,linecolor=blue](2.5,-0.5)(7.5,4.5)
\psline[linewidth=2pt,linecolor=blue](4.5,-0.5)(7.5,2.5)
\psline[linewidth=2pt,linecolor=blue](6.5,-0.5)(7.5,0.5)
\psline[linewidth=2pt,linecolor=blue](-0.5,0.5)(4.5,5.5)
\psline[linewidth=2pt,linecolor=blue](-0.5,2.5)(2.5,5.5)
\psline[linewidth=2pt,linecolor=blue](-0.5,4.5)(0.5,5.5)
\multirput{0}(0,0)(0,2){3}{%
  \multirput{0}(1,0)(2,0){4}{%
     \pscircle*[linecolor=white]{4pt}
     \pscircle[linewidth=1pt,linecolor=black] {4pt}
   }
}
\multirput{0}(0,1)(0,2){3}{%
  \multirput{0}(0,0)(1,0){8}{%
     \pscircle*[linecolor=white]{4pt}
     \pscircle[linewidth=1pt,linecolor=black] {4pt}
   }
}
\multirput{0}(0,0)(0,2){3}{%
  \multirput{0}(0,0)(2,0){4}{%
     \pscircle*{3pt}
   }
}
\endpspicture
\caption{ \label{fig.T34}
Kagom\'e lattice $T'(4,3)$: the vertex set is given by the white circles
($\circ$) and the edge set is given by the thick lines. This graph is
the medial graph of the triangulation $T(4,3)$, with vertex set given by the
solid black dots ($\bullet$) and edge set given by the thin gray lines. 
}
\end{figure}

An {\em edge-colouring} \footnote{The usual definition of edge-colourings  
is by colouring the edges in such a way that edges incident to
the same vertex receive distinct colours. 
In our definition, this actually works for the dual graph of the
triangulation. Another interpretation is to view edge-colourings as
vertex 3-colourings of the medial graph $M(G)$, which we shall do in the sequel. 
}
is a partition of the edges of a triangulation $T$
intro three classes, so that each triangular face of $T$ has one edge
in each class. This is equivalent to a proper three-colouring on the medial
graph $T'=M(T)$ of the triangulation $T$. 
(The precise definition of the medial graph is given below.) 
In particular, if $T=T(3L,3M)$,
then $T' = T'(3L,3M)$ is a {\em kagom\'e graph\/} embedded on a torus. On
Figure~\ref{fig.T34} we show the particular case of $T'(4,3)$. 

%
%
%
%

As mentioned above, there is a hierarchy among these types of colourings,
see \cite[Proposition 25]{Fisk_77b}. The simplest case of this hierarchy 
is the following one.

\begin{proposition}
\label{prop.Fisk.hierarchy}
Let\/ $T$ be a triangulation of a surface. Every four-colouring of\/ $T$
induces an edge-colouring of\/ $T$, and this correspondence is $1$-$1$.
\end{proposition}

\medskip

\noindent
{\bf Remarks.} 1. The reverse implication is false. Most triangulations have
edge-colourings that are not induced by any four-colouring. However, the two 
notions are equivalent for triangulations of the sphere. 

2. It is usually assumed that the zero-temperature triangular-lattice
4--state Potts antiferromagnet is {\em equivalent} to the zero-temperature
kagom\'e lattice 3-state Potts antiferromagnet. This is {\em not} true on
any surface other than the sphere, as there might be edge-colourings not
induced by four-colourings. See an example below. 

%
%
\begin{figure}[htb]
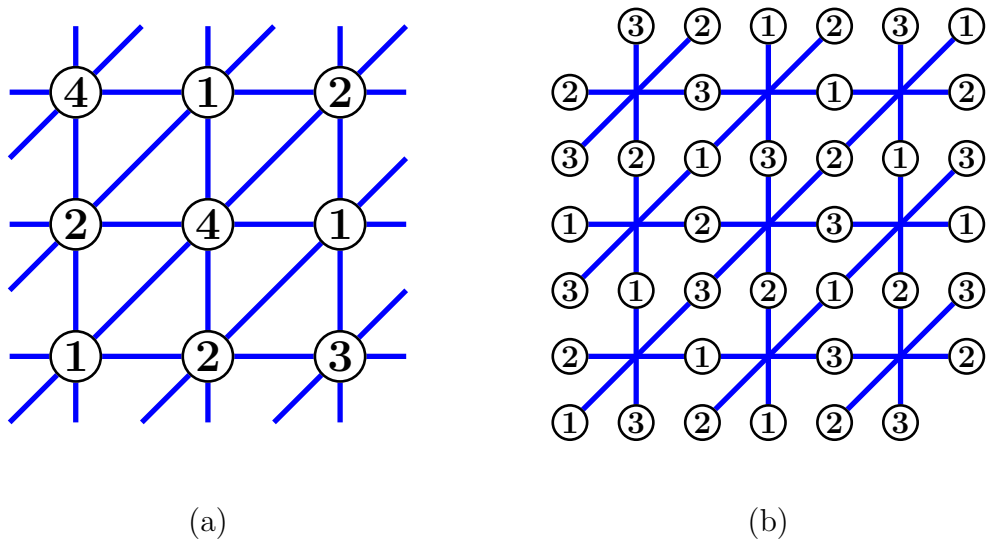

\centering
\begin{tabular}{cc}
%
%
\psset{xunit=50pt}
\psset{yunit=50pt}
\psset{labelsep=10pt}
\pspicture(-1,-1)(3,3)
\psline[linewidth=2pt,linecolor=blue](-0.5,0)(2.5,0)
\psline[linewidth=2pt,linecolor=blue](-0.5,1)(2.5,1)
\psline[linewidth=2pt,linecolor=blue](-0.5,2)(2.5,2)
\psline[linewidth=2pt,linecolor=blue](0,-0.5)(0,2.5)
\psline[linewidth=2pt,linecolor=blue](1,-0.5)(1,2.5)
\psline[linewidth=2pt,linecolor=blue](2,-0.5)(2,2.5)
\psline[linewidth=2pt,linecolor=blue](-0.5,-0.5)(2.5,2.5)
\psline[linewidth=2pt,linecolor=blue](0.5,-0.5)(2.5,1.5)
\psline[linewidth=2pt,linecolor=blue](1.5,-0.5)(2.5,0.5)
\psline[linewidth=2pt,linecolor=blue](-0.5,0.5)(1.5,2.5)
\psline[linewidth=2pt,linecolor=blue](-0.5,1.5)(0.5,2.5)
\multirput{0}(0,0)(0,1){3}{%
  \multirput{0}(0,0)(1,0){3}{%
     \pscircle*[linecolor=white]{10pt}
     \pscircle[linewidth=1pt,linecolor=black] {10pt}
   }
}
\rput{0}(0,0){{\Large \bf 1}}
\rput{0}(0,1){{\Large \bf 2}}
\rput{0}(0,2){{\Large \bf 4}}
\rput{0}(1,0){{\Large \bf 2}}
\rput{0}(1,1){{\Large \bf 4}}
\rput{0}(1,2){{\Large \bf 1}}
\rput{0}(2,0){{\Large \bf 3}}
\rput{0}(2,1){{\Large \bf 1}}
\rput{0}(2,2){{\Large \bf 2}}
\endpspicture
&
%
%
\psset{xunit=50pt}
\psset{yunit=50pt}
\psset{labelsep=10pt}
\pspicture(-1,-1)(3,3)
\psline[linewidth=2pt,linecolor=blue](-0.5,0)(2.5,0)
\psline[linewidth=2pt,linecolor=blue](-0.5,1)(2.5,1)
\psline[linewidth=2pt,linecolor=blue](-0.5,2)(2.5,2)
\psline[linewidth=2pt,linecolor=blue](0,-0.5)(0,2.5)
\psline[linewidth=2pt,linecolor=blue](1,-0.5)(1,2.5)
\psline[linewidth=2pt,linecolor=blue](2,-0.5)(2,2.5)
\psline[linewidth=2pt,linecolor=blue](-0.5,-0.5)(2.5,2.5)
\psline[linewidth=2pt,linecolor=blue](0.5,-0.5)(2.5,1.5)
\psline[linewidth=2pt,linecolor=blue](1.5,-0.5)(2.5,0.5)
\psline[linewidth=2pt,linecolor=blue](-0.5,0.5)(1.5,2.5)
\psline[linewidth=2pt,linecolor=blue](-0.5,1.5)(0.5,2.5)
\multirput{0}(-0.5,-0.5)(0,1){3}{%
  \multirput{0}(0,0)(1,0){3}{%
     \pscircle*[linecolor=white]{7pt}
     \pscircle[linewidth=1pt,linecolor=black] {7pt}
   }
}
\multirput{0}(-0.5,0)(0,1){3}{%
  \multirput{0}(0,0)(1,0){4}{%
     \pscircle*[linecolor=white]{7pt}
     \pscircle[linewidth=1pt,linecolor=black] {7pt}
   }
}
\multirput{0}(0,-0.5)(0,1){4}{%
  \multirput{0}(0,0)(1,0){3}{%
     \pscircle*[linecolor=white]{7pt}
     \pscircle[linewidth=1pt,linecolor=black] {7pt}
   }
}
\multirput{0}(2.5,0.5)(0,1){3}{
    \pscircle*[linecolor=white]{7pt}
    \pscircle[linewidth=1pt,linecolor=black] {7pt}
}
\multirput{0}(0.5,2.5)(1,0){2}{
    \pscircle*[linecolor=white]{7pt}
    \pscircle[linewidth=1pt,linecolor=black] {7pt}
}
\rput{0}(-0.5,-0.5){{\bf 1}}
\rput{0}( 0  ,-0.5){{\bf 3}}
\rput{0}( 0.5,-0.5){{\bf 2}}
\rput{0}( 1  ,-0.5){{\bf 1}}
\rput{0}( 1.5,-0.5){{\bf 2}}
\rput{0}( 2  ,-0.5){{\bf 3}}
\rput{0}(-0.5, 0.5){{\bf 3}}
\rput{0}( 0  , 0.5){{\bf 1}}
\rput{0}( 0.5, 0.5){{\bf 3}}
\rput{0}( 1  , 0.5){{\bf 2}}
\rput{0}( 1.5, 0.5){{\bf 1}}
\rput{0}( 2  , 0.5){{\bf 2}}
\rput{0}( 2.5, 0.5){{\bf 3}}
\rput{0}(-0.5, 1.5){{\bf 3}}
\rput{0}( 0  , 1.5){{\bf 2}}
\rput{0}( 0.5, 1.5){{\bf 1}}
\rput{0}( 1  , 1.5){{\bf 3}}
\rput{0}( 1.5, 1.5){{\bf 2}}
\rput{0}( 2  , 1.5){{\bf 1}}
\rput{0}( 2.5, 1.5){{\bf 3}}
\rput{0}( 0  , 2.5){{\bf 3}}
\rput{0}( 0.5, 2.5){{\bf 2}}
\rput{0}( 1  , 2.5){{\bf 1}}
\rput{0}( 1.5, 2.5){{\bf 2}}
\rput{0}( 2  , 2.5){{\bf 3}}
\rput{0}( 2.5, 2.5){{\bf 1}}
\rput{0}(-0.5, 0){{\bf 2}}
\rput{0}( 0.5, 0){{\bf 1}}
\rput{0}( 1.5, 0){{\bf 3}}
\rput{0}( 2.5, 0){{\bf 2}}
\rput{0}(-0.5, 1){{\bf 1}}
\rput{0}( 0.5, 1){{\bf 2}}
\rput{0}( 1.5, 1){{\bf 3}}
\rput{0}( 2.5, 1){{\bf 1}}
\rput{0}(-0.5, 2){{\bf 2}}
\rput{0}( 0.5, 2){{\bf 3}}
\rput{0}( 1.5, 2){{\bf 1}}
\rput{0}( 2.5, 2){{\bf 2}}
\endpspicture  \\
(a) & (b) 
\end{tabular}
\caption{\label{figure_tri_L=3A} 
(a) A four-colouring $f$ of $T(3,3)$ with $\deg(f)=0$. 
(b) The edge-colouring induced by $f$. 
}
\end{figure}
%
%

Let us explain how to obtain the edge-colouring induced by a given four-colouring
of a triangulation $T$. We will illustrate the general ideas with an 
example displayed in Figure~\ref{figure_tri_L=3A}: In (a) we plot a particular  
four-colouring $f$ with $\deg(f)=0$ of $T=T(3,3)$. 
The edge-colouring $g$ induced by $f$ 
is depicted in Figure~\ref{figure_tri_L=3A}(b). It is obtained
as follows: For each edge of $T$, we assign to the edge colour
$1$ if its end vertices are coloured $12$ or $34$; the edge will get colour
$2$ if its end vertices are coloured $13$ or $24$; and the edge gets colour 
$3$ if its end vertices are coloured $14$ or $23$. 
The three vertices on any triangular face $t\in T$ are coloured differently 
by $f$, thus the stated procedure colours the edges of $t$ with three
distinct colours.

We define a {\em Kempe region\/} for an edge-colouring \cite{Fisk_77b} 
in a similar fashion as for four-colourings. This is a region $R$ of the 
triangulation $T$ whose boundary has all edges of the same colour $c$. 
Then we can {\em exchange\/} the two colours different from $c$ on all edges 
in $R$.
Two edge-colourings are {\em K-equivalent\/} if one can be obtained from
the other by a sequence of exchanges on Kempe regions.

\begin{proposition}
\label{prop:Keq_induced}
If two four-colourings of a triangulation\/ $G$ are K-equivalent, 
then their induced edge-colourings are also K-equivalent.
\end{proposition}

\proof
Since the K-equivalence of four-colourings of $G$ is generated by K-exchanges on
the regions of the triangulation, it suffices to prove that a K-exchange for
a four-colouring $f$ made on a region $R$ corresponds to the exchange performed
on the same region for the induced edge-colouring $\psi$. Indeed, since $R$ is 
a Kempe region for $f$, all edges on the boundary of $R$ have their vertices 
coloured by the same pair of colours, say $a,b$. This implies that these edges 
have the same colour $c$ under the edge-colouring $\psi$. Now, exchanging 
the two colours different from $a,b$ on the vertices in the region $R$ has 
the effect on the induced edge-colouring that is precisely the same as the 
exchange of edge-colours different from $c$ on the edges in $R$. 
This completes the proof.  \hfill \qed 

\bigskip

\begin{figure}[ht]
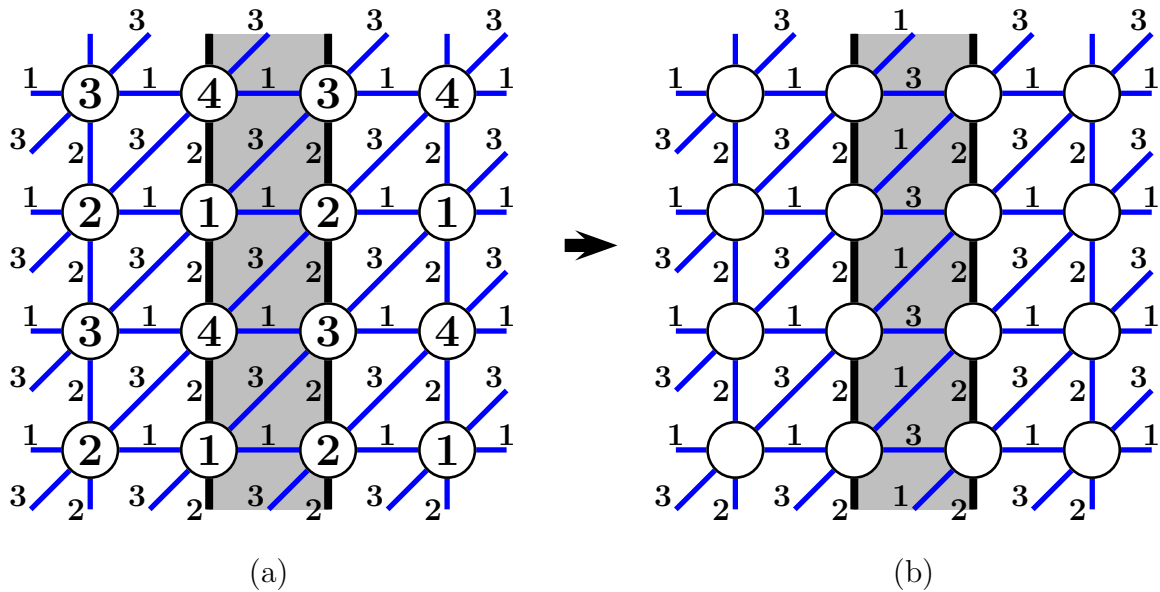

\centering
\begin{tabular}{ccc}
%
%
\psset{xunit=45pt}
\psset{yunit=45pt}
\psset{labelsep=2pt}
\pspicture(-0.5,-0.5)(3.5,3.5)
\psline*[linecolor=lightgray](1,-0.5)(1,3.5)(2,3.5)(2,-0.5)(1,-0.5)
\psline[linewidth=2pt,linecolor=blue](-0.5,0)(3.5,0)
\psline[linewidth=2pt,linecolor=blue](-0.5,1)(3.5,1)
\psline[linewidth=2pt,linecolor=blue](-0.5,2)(3.5,2)
\psline[linewidth=2pt,linecolor=blue](-0.5,3)(3.5,3)
\psline[linewidth=2pt,linecolor=blue](0,-0.5)(0,3.5)
\psline[linewidth=2pt,linecolor=blue](1,-0.5)(1,3.5)
\psline[linewidth=2pt,linecolor=blue](2,-0.5)(2,3.5)
\psline[linewidth=2pt,linecolor=blue](3,-0.5)(3,3.5)
\psline[linewidth=2pt,linecolor=blue](-0.5,-0.5)(3.5,3.5)
\psline[linewidth=2pt,linecolor=blue](0.5,-0.5)(3.5,2.5)
\psline[linewidth=2pt,linecolor=blue](1.5,-0.5)(3.5,1.5)
\psline[linewidth=2pt,linecolor=blue](2.5,-0.5)(3.5,0.5)
\psline[linewidth=2pt,linecolor=blue](-0.5,0.5)(2.5,3.5)
\psline[linewidth=2pt,linecolor=blue](-0.5,1.5)(1.5,3.5)
\psline[linewidth=2pt,linecolor=blue](-0.5,2.5)(0.5,3.5)
\psline[linewidth=3pt](1,-0.5)(1,3.5)
\psline[linewidth=3pt](2,-0.5)(2,3.5)
\multirput{0}(0,0)(0,1){4}{%
  \multirput{0}(0,0)(1,0){4}{%
     \pscircle*[linecolor=white]{11pt}
     \pscircle[linewidth=1pt,linecolor=black] {11pt}
   }
}
\rput{0}(0,0){{\Large \bf 2}}
\rput{0}(0,1){{\Large \bf 3}}
\rput{0}(0,2){{\Large \bf 2}}
\rput{0}(0,3){{\Large \bf 3}}
\rput{0}(1,0){{\Large \bf 1}}
\rput{0}(1,1){{\Large \bf 4}}
\rput{0}(1,2){{\Large \bf 1}}
\rput{0}(1,3){{\Large \bf 4}}
\rput{0}(2,0){{\Large \bf 2}}
\rput{0}(2,1){{\Large \bf 3}}
\rput{0}(2,2){{\Large \bf 2}}
\rput{0}(2,3){{\Large \bf 3}}
\rput{0}(3,0){{\Large \bf 1}}
\rput{0}(3,1){{\Large \bf 4}}
\rput{0}(3,2){{\Large \bf 1}}
\rput{0}(3,3){{\Large \bf 4}}
\multirput{0}(-1,0)(0,1){4}{%
  \multirput{0}(0,0)(1,0){5}{ \uput[90](0.5,0){{\bf 1}} }
}
\multirput{0}(-1,0)(0,1){3}{%
  \multirput{0}(0,0)(1,0){5}{ \uput[135](0.5,0.5){{\bf 3}} }
}
\multirput{0}(0,3)(1,0){4}{ \uput[135](0.5,0.5){{\bf 3}} }
\multirput{0}(-1,-1)(1,0){4}{ \uput[135](0.5,0.5){{\bf 3}} }
\multirput{0}(0,-1)(0,1){4}{%
  \multirput{0}(0,0)(1,0){4}{ \uput[180](0,0.5){{\bf 2}} }
}
\endpspicture
&
\psset{xunit=50pt}
\psset{yunit=50pt}
\pspicture(-0.4,-0.5)(0.4,3.5)
\psline[linewidth=5pt]{->}(-0.2,1.5)(0.2,1.5)
\endpspicture
& 
%
%
\psset{xunit=45pt}
\psset{yunit=45pt}
\psset{labelsep=2pt}
\pspicture(-0.5,-0.5)(3.5,3.5)
\psline*[linecolor=lightgray](1,-0.5)(1,3.5)(2,3.5)(2,-0.5)(1,-0.5)
\psline[linewidth=2pt,linecolor=blue](-0.5,0)(3.5,0)
\psline[linewidth=2pt,linecolor=blue](-0.5,1)(3.5,1)
\psline[linewidth=2pt,linecolor=blue](-0.5,2)(3.5,2)
\psline[linewidth=2pt,linecolor=blue](-0.5,3)(3.5,3)
\psline[linewidth=2pt,linecolor=blue](0,-0.5)(0,3.5)
\psline[linewidth=2pt,linecolor=blue](1,-0.5)(1,3.5)
\psline[linewidth=2pt,linecolor=blue](2,-0.5)(2,3.5)
\psline[linewidth=2pt,linecolor=blue](3,-0.5)(3,3.5)
\psline[linewidth=2pt,linecolor=blue](-0.5,-0.5)(3.5,3.5)
\psline[linewidth=2pt,linecolor=blue](0.5,-0.5)(3.5,2.5)
\psline[linewidth=2pt,linecolor=blue](1.5,-0.5)(3.5,1.5)
\psline[linewidth=2pt,linecolor=blue](2.5,-0.5)(3.5,0.5)
\psline[linewidth=2pt,linecolor=blue](-0.5,0.5)(2.5,3.5)
\psline[linewidth=2pt,linecolor=blue](-0.5,1.5)(1.5,3.5)
\psline[linewidth=2pt,linecolor=blue](-0.5,2.5)(0.5,3.5)
\psline[linewidth=3pt](1,-0.5)(1,3.5)
\psline[linewidth=3pt](2,-0.5)(2,3.5)
\multirput{0}(0,0)(0,1){4}{%
  \multirput{0}(0,0)(1,0){4}{%
     \pscircle*[linecolor=white]{11pt}
     \pscircle[linewidth=1pt,linecolor=black] {11pt}
   }
}
\multirput{0}(-1,0)(0,1){4}{ \uput[90](0.5,0){{\bf 1}}}
\multirput{0}( 0,0)(0,1){4}{ \uput[90](0.5,0){{\bf 1}}}
\multirput{0}( 1,0)(0,1){4}{ \uput[90](0.5,0){{\bf 3}}}
\multirput{0}( 2,0)(0,1){4}{ \uput[90](0.5,0){{\bf 1}}}
\multirput{0}( 3,0)(0,1){4}{ \uput[90](0.5,0){{\bf 1}}}
\multirput{0}(-1,-1)(0,1){4}{ \uput[135](0.5,0.5){{\bf 3}}}
\multirput{0}( 0,-1)(0,1){5}{ \uput[135](0.5,0.5){{\bf 3}}}
\multirput{0}( 1,-1)(0,1){5}{ \uput[135](0.5,0.5){{\bf 1}}}
\multirput{0}( 2,-1)(0,1){5}{ \uput[135](0.5,0.5){{\bf 3}}}
\multirput{0}( 3, 0)(0,1){4}{ \uput[135](0.5,0.5){{\bf 3}}}
\multirput{0}(0,-1)(0,1){4}{%
  \multirput{0}(0,0)(1,0){4}{ \uput[180](0,0.5){{\bf 2}} }
}
\endpspicture
\\[4mm]
(a) &  & (b)
\end{tabular}
\caption{ \label{fig.newfigure2}
Example showing that the converse of Proposition~\ref{prop:Keq_induced}
cannot be true. In panel (a) we show a 4-colouring of the triangulation
$T(4,4)$ and its associated edge-colouring. The shaded triangles form
a Kempe region for this edge-colouring. In panel~(b), we display the
new edge-colouring after the Kempe change. The exchange yields an
edge-colouring that is not induced by any four-colouring. This example can
be generalized to any triangulation $T(2n,2m)$, with $n,m\ge 2$.  
}
\end{figure}

It is natural to ask if a converse of Proposition~\ref{prop:Keq_induced}
may hold. Unfortunately, the answer is negative. Even more, an edge-colouring
$\psi$ induced by a four-colouring $f$ may be K-equivalent to an edge-colouring
that is not induced by any four-colouring. 
Figure~\ref{fig.newfigure2} shows an example of such a case.

%
%
%
%

Although there are at least as many
edge-colourings as there are four-colourings, the number of K-equivalence 
classes of four-colourings might be bigger or might be smaller than the
number of equivalence classes of edge-colourings. 

Our analysis can be simplified by introducing
the {\em medial graph\/} $T'=M(T)= (V',E')$ of a triangulation $T$.

\begin{figure}[htb]
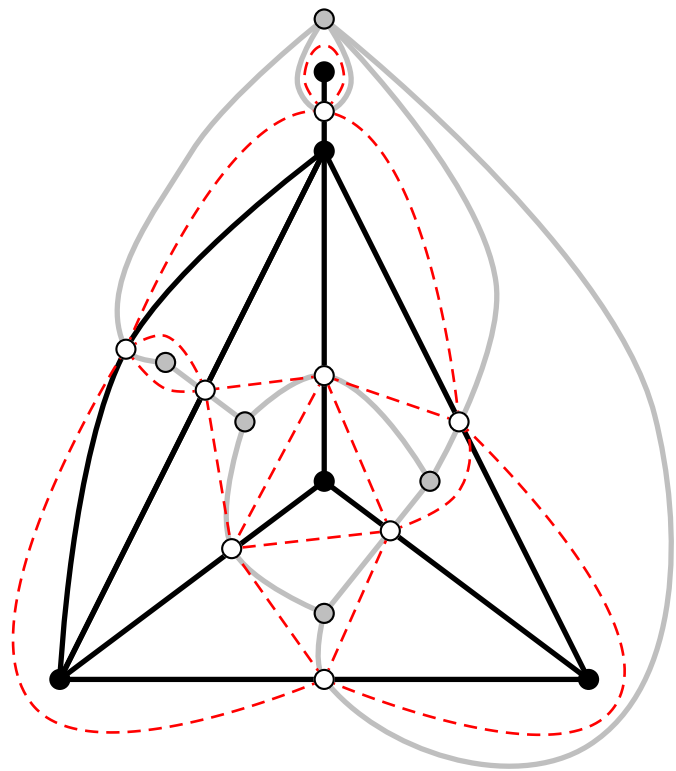

\centering
\psset{xunit=50pt}
\psset{yunit=50pt}
\psset{labelsep=5pt}
\pspicture(0,0)(6,6)
\psline[linewidth=2pt]{cc-cc}(1,1)(5,1)(3,5)(1,1)
\psline[linewidth=2pt]{cc-cc}(1,1)(3,2.5)(3,5)
\psline[linewidth=2pt]{cc-cc}(3,2.5)(5,1)
\psline[linewidth=2pt]{cc-cc}(1,1)(3,5)(3,5.6)
\pscurve[linewidth=2pt]{cc-cc}(1,1)(1.5,3.5)(3,5)

\psline[linewidth=2pt,linecolor=lightgray](3,1.5)(3.8,2.5)
\psline[linewidth=2pt,linecolor=lightgray](2.4,2.95)(1.8,3.4)
\pscurve[linewidth=2pt,linecolor=lightgray](2.4,2.95)(2.3,2)(3,1.5)
\pscurve[linewidth=2pt,linecolor=lightgray](2.4,2.95)(3,3.3)(3.8,2.5)
\pscurve[linewidth=2pt,linecolor=lightgray](1.8,3.4)(1.5,3.5)(2,5)(3,6)
\pscurve[linewidth=2pt,linecolor=lightgray](3.8,2.5)(4.3,4)(3,6)
\pscurve[linewidth=2pt,linecolor=lightgray](3,1.5)(3,1)(5,0.5)(5.5,3)(3,6)
\pscurve[linewidth=2pt,linecolor=lightgray](3,6)(2.8,5.5)(3,5.3)(3.2,5.5)(3,6)

\psline[linewidth=1pt,linestyle=dashed,linecolor=red](3,1)(2.3,1.99)%
(3.5,2.125)(3,1)
\psline[linewidth=1pt,linestyle=dashed,linecolor=red](3.5,2.125)(3,3.3)%
(4.02,2.95)
\pscurve[linewidth=1pt,linestyle=dashed,linecolor=red](3.5,2.125)(4,2.4)%
(4.1,2.8)(4.02,2.95)
\psline[linewidth=1pt,linestyle=dashed,linecolor=red](3,3.3)(2.1,3.19)%
(2.3,1.99)(3,3.3)
\pscurve[linewidth=1pt,linestyle=dashed,linecolor=red](1.5,3.5)(1.8,3.6)%
(2.1,3.19)
\pscurve[linewidth=1pt,linestyle=dashed,linecolor=red](1.5,3.5)(1.8,3.2)%
(2.1,3.19)
\pscurve[linewidth=1pt,linestyle=dashed,linecolor=red](1.5,3.5)(3,5.3)%
(4.02,2.95)
\pscurve[linewidth=1pt,linestyle=dashed,linecolor=red](4.02,2.95)(5.2,0.8)(3,1)
\pscurve[linewidth=1pt,linestyle=dashed,linecolor=red](3,1)(0.8,0.8)(1.5,3.5)
\pscurve[linewidth=1pt,linestyle=dashed,linecolor=red](3,5.3)(2.85,5.55)%
(3,5.8)(3.15,5.55)(3,5.3)

\pscircle*[linecolor=white](3,1){4pt}
\pscircle(3,1){4pt}
\pscircle*[linecolor=white](3.5,2.125){4pt}
\pscircle(3.5,2.125){4pt}
\pscircle*[linecolor=white](2.1,3.19){4pt}
\pscircle(2.1,3.19){4pt}
\pscircle*[linecolor=white](1.5,3.5){4pt}
\pscircle(1.5,3.5){4pt}
\pscircle*[linecolor=white](4.02,2.95){4pt}
\pscircle(4.02,2.95){4pt}
\pscircle*[linecolor=white](3,3.3){4pt}
\pscircle(3,3.3){4pt}
\pscircle*[linecolor=white](2.3,1.99){4pt}
\pscircle(2.3,1.99){4pt}
\pscircle*[linecolor=white](3,5.3){4pt}
\pscircle(3,5.3){4pt}

\pscircle*[linecolor=lightgray](3,1.5){4pt}
\pscircle(3,1.5){4pt}
\pscircle*[linecolor=lightgray](3.8,2.5){4pt}
\pscircle(3.8,2.5){4pt}
\pscircle*[linecolor=lightgray](3,6){4pt}
\pscircle(3,6){4pt}
\pscircle*[linecolor=lightgray](2.4,2.95){4pt}
\pscircle(2.4,2.95){4pt}
\pscircle*[linecolor=lightgray](1.8,3.4){4pt}
\pscircle(1.8,3.4){4pt}

\pscircle*(1,1)   {4pt}
\pscircle*(5,1)   {4pt}
\pscircle*(3,5)   {4pt}
\pscircle*(3,2.5) {4pt}
\pscircle*(3,5.6) {4pt}
\endpspicture
\caption{\label{figure_medial}
Graph $G=(V,E)$ embedded on the sphere. The vertices of $V$ are depicted
as solid black circles, and the edges of $E$ as solid thick lines. 
The dual graph $G^*=(V^*,E^*)$ is represented as follows: the vertex set is 
depicted as solid gray circles, and the edge set as solid thick gray lines. 
Finally, the medial graph $G'=(V',E')$ is given by a vertex set drawn as
open white circles, and by the edge set depicted as dashed thin (red) lines.  
}
\end{figure}

Let us first define the medial graph $G'$ of a graph $G$ 
(not necessarily a triangulation) embedded on a surface $S$.
It is convenient to first define the {\em dual graph\/} $G^*=(V^*,E^*)$ 
of $G$, which is also embedded on $S$. 
This dual graph is built in the standard way as follows:
To each face $f$ in $G$, there corresponds a dual vertex $f^*\in V^*$;
and for every edge $e\in E$, we draw a dual edge $e^* \in E^*$. If the
original edge $e$ lies on the intersection of two faces $f$ and $h$
(possibly $f=h$), then the corresponding dual edge $e^*$ joins the
dual vertices $f^*,h^* \in V^*$. We can draw $G$ and $G^*$ on $S$ in such
a way that each edge $e\in E$ intersects its corresponding dual edge 
$e^* \in E^*$ exactly once.  
(See Figure~\ref{figure_medial} for an example on the sphere.)

%
%
%
%

Then, the {\em medial graph\/} $G'=(V',E')$ of $G=(V,E)$ 
is constructed as follows: 
To each unique intersection between an edge $e\in E$ and its dual edge
$e^*\in E^*$,  there corresponds a vertex of the medial graph $v' \in V'$. 
(See Figure~\ref{figure_medial} where the vertices of the medial graph are
depicted as open white circles.) 
It is clear that the medial graph $G'$ is
also embedded on $S$, and that $G'$ is a regular graph of degree 4.  
Finally, the role played in this construction by $G$ and its dual $G^*$ is 
symmetric; therefore, the medial graph of $G$ coincides with
the medial of its dual $(G^*)' = G'$. 

\begin{figure}[htb]
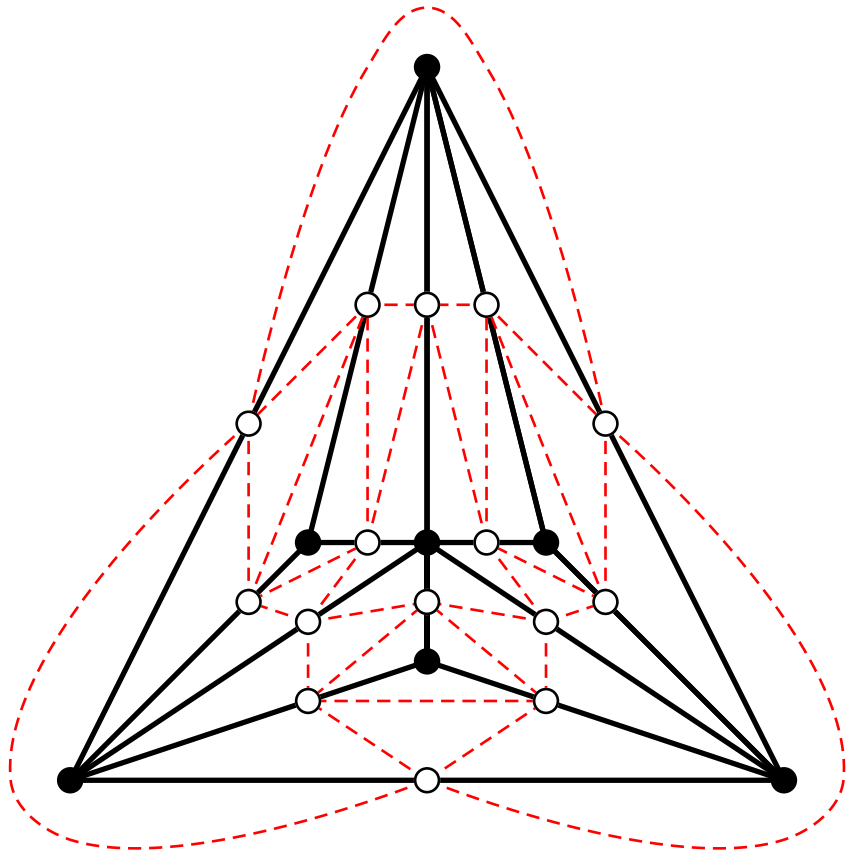

\centering
\psset{xunit=45pt}
\psset{yunit=45pt}
\psset{labelsep=5pt}
\pspicture(-1,-1)(7,7)
\psline[linewidth=2pt]{cc-cc}(0,0)(6,0)(3,6)(0,0)%
\psline[linewidth=2pt]{cc-cc}(0,0)(2,2)(3,6)%
\psline[linewidth=2pt]{cc-cc}(3,6)(3,2)(2,2)%
\psline[linewidth=2pt]{cc-cc}(6,0)(4,2)(3,6)%
\psline[linewidth=2pt]{cc-cc}(6,0)(4,2)(3,6)%
\psline[linewidth=2pt]{cc-cc}(3,6)(3,2)(4,2)%
\psline[linewidth=2pt]{cc-cc}(0,0)(3,2)(6,0)%
\psline[linewidth=2pt]{cc-cc}(0,0)(3,1)(3,2)(3,1)(6,0)%
\psline[linewidth=1pt,linecolor=red,linestyle=dashed](2,1.333)(2,0.667)%
(3,1.5)(2,1.333)
\psline[linewidth=1pt,linecolor=red,linestyle=dashed](4,1.333)(4,0.667)%
(3,1.5)(4,1.333)
\psline[linewidth=1pt,linecolor=red,linestyle=dashed](2,1.333)(2.5,2)(1.5,1.5)%
(2,1.333)
\psline[linewidth=1pt,linecolor=red,linestyle=dashed](4,1.333)(3.5,2)(4.5,1.5)%
(4,1.333)
\psline[linewidth=1pt,linecolor=red,linestyle=dashed](1.5,1.5)(1.5,3)(2.5,4)%
(1.5,1.5)
\psline[linewidth=1pt,linecolor=red,linestyle=dashed](4.5,1.5)(4.5,3)(3.5,4)%
(4.5,1.5)
\psline[linewidth=1pt,linecolor=red,linestyle=dashed](3,0)(4,0.667)(2,0.667)%
(3,0)
\psline[linewidth=1pt,linecolor=red,linestyle=dashed](2.5,2)(2.5,4)(3,4)(2.5,2)
\psline[linewidth=1pt,linecolor=red,linestyle=dashed](3.5,2)(3.5,4)(3,4)(3.5,2)
\pscurve[linewidth=1pt,linecolor=red,linestyle=dashed](3,0)(6,-0.5)(6.5,0)%
(4.5,3)
\pscurve[linewidth=1pt,linecolor=red,linestyle=dashed](3,0)(0,-0.5)(-0.5,0)%
(1.5,3)
\pscurve[linewidth=1pt,linecolor=red,linestyle=dashed](1.5,3)(2.5,6)(3,6.5)%
(3.5,6)(4.5,3)

\rput{0}(0,0){\pscircle*{5pt}}
\rput{0}(6,0){\pscircle*{5pt}}
\rput{0}(2,2){\pscircle*{5pt}}
\rput{0}(3,2){\pscircle*{5pt}}
\rput{0}(4,2){\pscircle*{5pt}}
\rput{0}(3,6){\pscircle*{5pt}}
\rput{0}(3,1){\pscircle*{5pt}}
\rput{0}(1.5,1.5){\pscircle*[linecolor=white]{5pt}
                  \pscircle[linewidth=1pt]{5pt}
}
\rput{0}(4.5,1.5){\pscircle*[linecolor=white]{5pt}
                  \pscircle[linewidth=1pt]{5pt}
}
\rput{0}(2.5,2){\pscircle*[linecolor=white]{5pt}
                \pscircle[linewidth=1pt]{5pt}
}
\rput{0}(3.5,2){\pscircle*[linecolor=white]{5pt}
                \pscircle[linewidth=1pt]{5pt}
}
\rput{0}(3,1.5){\pscircle*[linecolor=white]{5pt}
                \pscircle[linewidth=1pt]{5pt}
}
\rput{0}(2.5,4){\pscircle*[linecolor=white]{5pt}
                \pscircle[linewidth=1pt]{5pt}
}
\rput{0}(3.5,4){\pscircle*[linecolor=white]{5pt}
                \pscircle[linewidth=1pt]{5pt}
}
\rput{0}(3,4){\pscircle*[linecolor=white]{5pt}
              \pscircle[linewidth=1pt]{5pt}
}
\rput{0}(2,0.667){\pscircle*[linecolor=white]{5pt}
                  \pscircle[linewidth=1pt]{5pt}
}
\rput{0}(4,0.667){\pscircle*[linecolor=white]{5pt}
                  \pscircle[linewidth=1pt]{5pt}
}
\rput{0}(4,1.333){\pscircle*[linecolor=white]{5pt}
                  \pscircle[linewidth=1pt]{5pt}
}
\rput{0}(2,1.333){\pscircle*[linecolor=white]{5pt}
                  \pscircle[linewidth=1pt]{5pt}
}
\rput{0}(1.5,3){\pscircle*[linecolor=white]{5pt}
                \pscircle[linewidth=1pt]{5pt}
}
\rput{0}(4.5,3){\pscircle*[linecolor=white]{5pt}
                \pscircle[linewidth=1pt]{5pt}
}
\rput{0}(3,0){\pscircle*[linecolor=white]{5pt}
              \pscircle[linewidth=1pt]{5pt}
}
\endpspicture
\caption{ \label{figure_medial_tri}
Triangulation $T=(V,E)$ of the sphere. The vertices of $V$ are depicted
as solid black circles, and the edges of $E$ as solid thick lines. 
Its medial graph $T'=(V',E')$ is represented as follows: the vertices 
are depicted as open white circles, and the edges as dashed thin lines.  
The triangulation $T''=(V'',E'')$ is constructed as follows: 
the vertices in $V''$
are those of $V$ and $V'$ (i.e., all circles, open or solid in the 
picture). The edges in $E''$ are all edges in the picture (dashed and solid);
notice that each edge in $E$ now corresponds to two edges in $E''$.  
For simplicity, we have not depicted $T^*$, the dual graph of $T$.
}
\end{figure}

Edge-colourings of a triangulation $T=(V,E)$ embedded on 
a surface $S$ can be regarded as three-colourings of the 
vertices of the corresponding medial graph $T'=M(T)$, which is also
embedded on $S$. 
In $T'$ there are two types of faces: triangular faces inside any 
triangular face of $T$, and faces with $d_i$ sides containing every 
vertex $i\in V$ of degree $d_i$. (See Figure~\ref{figure_medial_tri}
for an example of a triangulation embedded on the sphere.) 
Notice that the medial graph $T'$ of a triangulation $T$ is not 
a triangulation of $S$ (with the exception when $T=K_4$). 
A Kempe change on an edge-colouring of $T$ has precisely the same effect
as a standard Kempe change on $T'$. 
 
%
%
%
%

The medial graph $T'=(V',E')$ can be regarded as a particular subgraph 
of another triangulation $T''=(V'',E'')$ on the same surface that is obtained
by adding back the original vertices of $T$, and 
also adding the edges joining each $v\in V$
with the vertices in $V'$ corresponding to the edges of $T$ incident with $v$.
The vertices of $T''$ are simply $V''=V \cup V'$. 
The edges of $T''$ are given by 
$E'' = E' \cup \widetilde{E}$. The second edge set $\widetilde{E}$
is constructed as follows: 
To each original edge $ab\in E$, the medial graph $T'$ assigns a 
new vertex $x$. Then, the contribution of the edge $ab\in E$ to 
$\widetilde{E}$ consists of two edges $ax$ and $xb$. 
Notice that the degree in $T''$ of the vertices $i$
in $V$ (resp.\ $V'$) is $d_i$ (resp.\ six).  
The triangulation $T''$ can be constructed directly from $T$ by inserting 
a new vertex in the middle of each edge of $T$ and then inserting three edges 
joining the added vertices in each triangular face. Consequently, each 
face of $T$ is subdivided into four triangles in $T''$.  
Clearly, the resulting triangulation $T''$ triangulates the same surface $S$. 
(See Figure~\ref{figure_medial_tri}.)

Any 3-colouring of the medial graph $T'$ can be
regarded as a very particular four-colouring of $T''$, in which every
original vertex $i\in V$ (of degree $d_i$) is coloured $4$, and the vertices 
in $V'$ (of degree six in $T''$) are coloured $1,2,3$ in such a way the 
resulting 
four-colouring is proper. The Kempe changes on $T'$ can be viewed as a subset 
of the full set of possible Kempe changes we can perform on $T''$: 
In particular, we can only choose the induced subgraphs $T''_{ab}$ with
$a,b\in \{1,2,3\}$.  

We can regard the edge-colourings of a triangulation $T$ as constrained
colourings on $T''$; and we can use all the technology we have for
standard four-colourings of triangulations (in particular, the notion 
of the colouring degree) \cite{Mohar_Salas}. 

\begin{figure}[htb]
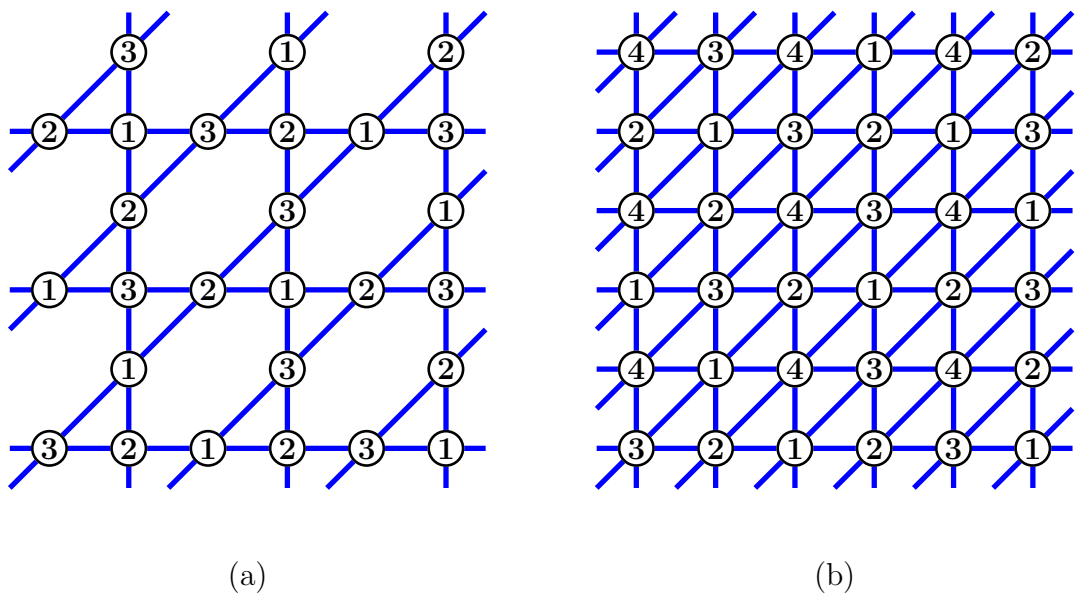

\centering
\begin{tabular}{cc}
%
%
\psset{xunit=30pt}
\psset{yunit=30pt}
\pspicture(-1,-1)(6,6)
\psline[linewidth=2pt,linecolor=blue](-0.5,0)(5.5,0)
\psline[linewidth=2pt,linecolor=blue](-0.5,2)(5.5,2)
\psline[linewidth=2pt,linecolor=blue](-0.5,4)(5.5,4)
\psline[linewidth=2pt,linecolor=blue](1,-0.5)(1,5.5)
\psline[linewidth=2pt,linecolor=blue](3,-0.5)(3,5.5)
\psline[linewidth=2pt,linecolor=blue](5,-0.5)(5,5.5)
\psline[linewidth=2pt,linecolor=blue](-0.5,-0.5)(5.5,5.5)
\psline[linewidth=2pt,linecolor=blue](1.5,-0.5)(5.5,3.5)
\psline[linewidth=2pt,linecolor=blue](3.5,-0.5)(5.5,1.5)
\psline[linewidth=2pt,linecolor=blue](-0.5,1.5)(3.5,5.5)
\psline[linewidth=2pt,linecolor=blue](-0.5,3.5)(1.5,5.5)
\multirput{0}(0,0)(0,2){3}{%
  \multirput{0}(0,0)(1,0){6}{%
     \pscircle*[linecolor=white]{7pt}
     \pscircle[linewidth=1pt,linecolor=black] {7pt}
   }
}
\multirput{0}(1,1)(0,2){3}{%
  \multirput{0}(0,0)(2,0){3}{%
     \pscircle*[linecolor=white]{7pt}
     \pscircle[linewidth=1pt,linecolor=black] {7pt}
   }
}

\rput{0}(0,0){\bf 3}
\rput{0}(0,2){\bf 1}
\rput{0}(0,4){\bf 2}

\rput{0}(1,0){\bf 2}
\rput{0}(1,1){\bf 1}
\rput{0}(1,2){\bf 3}
\rput{0}(1,3){\bf 2}
\rput{0}(1,4){\bf 1}
\rput{0}(1,5){\bf 3}

\rput{0}(2,0){\bf 1}
\rput{0}(2,2){\bf 2}
\rput{0}(2,4){\bf 3}

\rput{0}(3,0){\bf 2}
\rput{0}(3,1){\bf 3}
\rput{0}(3,2){\bf 1}
\rput{0}(3,3){\bf 3}
\rput{0}(3,4){\bf 2}
\rput{0}(3,5){\bf 1}

\rput{0}(4,0){\bf 3}
\rput{0}(4,2){\bf 2}
\rput{0}(4,4){\bf 1}

\rput{0}(5,0){\bf 1}
\rput{0}(5,1){\bf 2}
\rput{0}(5,2){\bf 3}
\rput{0}(5,3){\bf 1}
\rput{0}(5,4){\bf 3}
\rput{0}(5,5){\bf 2}
\endpspicture
&
%
%
\psset{xunit=30pt}
\psset{yunit=30pt}
\pspicture(-1,-1)(6,6)
\psline[linewidth=2pt,linecolor=blue](-0.5,0)(5.5,0)
\psline[linewidth=2pt,linecolor=blue](-0.5,1)(5.5,1)
\psline[linewidth=2pt,linecolor=blue](-0.5,2)(5.5,2)
\psline[linewidth=2pt,linecolor=blue](-0.5,3)(5.5,3)
\psline[linewidth=2pt,linecolor=blue](-0.5,4)(5.5,4)
\psline[linewidth=2pt,linecolor=blue](-0.5,5)(5.5,5)

\psline[linewidth=2pt,linecolor=blue](0,-0.5)(0,5.5)
\psline[linewidth=2pt,linecolor=blue](1,-0.5)(1,5.5)
\psline[linewidth=2pt,linecolor=blue](2,-0.5)(2,5.5)
\psline[linewidth=2pt,linecolor=blue](3,-0.5)(3,5.5)
\psline[linewidth=2pt,linecolor=blue](4,-0.5)(4,5.5)
\psline[linewidth=2pt,linecolor=blue](5,-0.5)(5,5.5)

\psline[linewidth=2pt,linecolor=blue](-0.5,-0.5)(5.5,5.5)
\psline[linewidth=2pt,linecolor=blue](0.5,-0.5)(5.5,4.5)
\psline[linewidth=2pt,linecolor=blue](1.5,-0.5)(5.5,3.5)
\psline[linewidth=2pt,linecolor=blue](2.5,-0.5)(5.5,2.5)
\psline[linewidth=2pt,linecolor=blue](3.5,-0.5)(5.5,1.5)
\psline[linewidth=2pt,linecolor=blue](4.5,-0.5)(5.5,0.5)
\psline[linewidth=2pt,linecolor=blue](-0.5,0.5)(4.5,5.5)
\psline[linewidth=2pt,linecolor=blue](-0.5,1.5)(3.5,5.5)
\psline[linewidth=2pt,linecolor=blue](-0.5,2.5)(2.5,5.5)
\psline[linewidth=2pt,linecolor=blue](-0.5,3.5)(1.5,5.5)
\psline[linewidth=2pt,linecolor=blue](-0.5,4.5)(0.5,5.5)
\multirput{0}(0,0)(0,1){6}{%
  \multirput{0}(0,0)(1,0){6}{%
     \pscircle*[linecolor=white]{7pt}
     \pscircle[linewidth=1pt,linecolor=black] {7pt}
   }
}

\rput{0}(0,0){\bf 3}
\rput{0}(0,1){\bf 4}
\rput{0}(0,2){\bf 1}
\rput{0}(0,3){\bf 4}
\rput{0}(0,4){\bf 2}
\rput{0}(0,5){\bf 4}

\rput{0}(1,0){\bf 2}
\rput{0}(1,1){\bf 1}
\rput{0}(1,2){\bf 3}
\rput{0}(1,3){\bf 2}
\rput{0}(1,4){\bf 1}
\rput{0}(1,5){\bf 3}

\rput{0}(2,0){\bf 1}
\rput{0}(2,1){\bf 4}
\rput{0}(2,2){\bf 2}
\rput{0}(2,3){\bf 4}
\rput{0}(2,4){\bf 3}
\rput{0}(2,5){\bf 4}

\rput{0}(3,0){\bf 2}
\rput{0}(3,1){\bf 3}
\rput{0}(3,2){\bf 1}
\rput{0}(3,3){\bf 3}
\rput{0}(3,4){\bf 2}
\rput{0}(3,5){\bf 1}

\rput{0}(4,0){\bf 3}
\rput{0}(4,1){\bf 4}
\rput{0}(4,2){\bf 2}
\rput{0}(4,3){\bf 4}
\rput{0}(4,4){\bf 1}
\rput{0}(4,5){\bf 4}

\rput{0}(5,0){\bf 1}
\rput{0}(5,1){\bf 2}
\rput{0}(5,2){\bf 3}
\rput{0}(5,3){\bf 1}
\rput{0}(5,4){\bf 3}
\rput{0}(5,5){\bf 2}
\endpspicture
\\[2mm]
   (a) & (b) \\[5mm]
\end{tabular}
\caption{\label{figure_tri_L=3B} 
Different colourings of the triangulation $T(3,3)$ and their interpretation. 
(a) The representation of the edge-colouring $g$ given in
Figure~\protect\ref{figure_tri_L=3A}(b) as a three-colouring $h$ of the
medial graph $T'(3,3)$, which is a kagom\'e graph embedded on a torus.
(b) The representation of the three-colouring $h$ on $T'(3,3)$ as 
a four-colouring of $T''(3,3)=T(6,6)$. 
[Notice that this panel is obtained from 
Figure~\protect\ref{figure_tri_L=3A}(b) by periodically shifting the 
later one step to the left, and by placing 
a spin coloured 4 on every intersection of the edges in 
Figure~\protect\ref{figure_tri_L=3A}(b).]
}
\end{figure}

These observations hold for every triangulation $T$. In this paper, 
we are focusing on the triangulations $T=T(3L,3M)$ of the torus. For these
particular triangulations, the corresponding medial graphs 
$T'=M(T)=T'(3L,3M)$ are kagom\'e graphs embedded on a torus. 
It is easy to see that the graph $T''$ is isomorphic to the triangulation
$T(6L,6M)$ in this case. 
(See Figures~\ref{figure_tri_L=3A} and~\ref{figure_tri_L=3B} 
for an explicit example with $L=M=1$.)

%
%

Given the triangulation $T=T(3L,3M)$, the set of proper four-colourings
on $T''=T(6L,6M)$ is denoted $\mathcal{C}_4(T'')$; the set of 
constrained proper four-colourings of $T$ with all vertices in 
$V$ coloured $4$ and those in $V'$ coloured $1,2,3$, will be denoted 
$\widetilde{\mathcal{C}}_4(T'')$.
The colourings in $\widetilde{\mathcal{C}}_4(T'')$ will be referred to as
the {\em special four-colourings\/} of $T''$.

Let us summarize the described correspondence in the following proposition:

\begin{proposition}
\label{prop:summarize}
Let us consider a triangulation $T$ embedded on a surface $S$. Then, 
there is a bijective correspondence between the three--edge-colourings of $T$ 
and the three-colourings of the vertices of its medial graph $T'=M(T)$.
Furthermore, if $T=T(3L,3M)$, then $T'$ is a subgraph of $T''=T(6L,6M)$ and 
there is a bijection between the three-colourings of the vertices of 
the medial graph $T'=M(T)$ and the special four-colourings of $T''$.
Under these two correspondences, K-equivalence is preserved. 
\end{proposition}

The main lemma we need is the following:

\begin{lemma} \label{lemma.kag}
Let us consider the set of edge-colourings of the triangulation $T(3L,3M)$. If
the set of special four-colourings of $T(6L,6M)$ contains an
element $f$ with $\deg(f)\equiv 6 \pmod{12}$, then there are at least
two Kempe equivalence classes for edge-colourings of $T(3L,3M)$. 
In other words, there are at least two Kempe equivalence classes for the 
vertex three-colourings of the medial graph $T'(3L,3M)$. 
\end{lemma}

\proof
First of all, let us prove that there is an element
$h \in \widetilde{\mathcal{C}}_4(T(6L,6M))$ with $\deg(h)=0$. This colouring 
will belong to one Kempe equivalence class for edge-colourings of $T(3L,3M)$. 
This four-colouring can be constructed as follows: We start with the
standard three-colouring of $T''=T(6L,6M)$, which always exists as 
both dimensions are multiple of three. Next, we change the colour of every 
vertex in $T''$ coming from the original triangulation $T=T(3L,3M)$ 
(viewed as a special vertex in $T''$) into the colour $4$. 
Recolouring of each such vertex is a Kempe change on $T''$,
so this gives a special colouring $h$ that is K-equivalent to the three-colouring
of $T''$. Since the degree of the three-colouring is 0 and changing the colour
of a single vertex preserves the degree of the colouring,
we conclude that $\deg(h)=0$.  

Suppose now that there is another proper four-colouring $f$ of $T(6L,6M)$
belonging to $\widetilde{\mathcal{C}}_4(T(6L,6M))$ and such that
$\deg(f)\equiv 6 \pmod{12}$. Theorem~3.5 of \cite{Mohar_Salas} 
ensures that the 
four-colourings $h$ and $f$ are not K-equivalent on the larger 
configuration space $\mathcal{C}_4(T(6L,6M))$. Thus, this conclusion holds
if we restrict to the smaller space 
$\widetilde{\mathcal{C}}_4(T(6L,6M))\subseteq \mathcal{C}_4(T(6L,6M))$
of special four-colourings. 
\mbox{} \hfill \qed

\bigskip

\noindent
{\bf Remark}. The proper 3--colouring $h$ described in the above proof for
$T'(3L,3L)$ corresponds to the so-called `$\sqrt{3}\times\sqrt{3}$' ordered
state in the physics literature \cite{Huse}.

%
%
\section{Main result} \label{sec.main} 

The goal of this section is to prove the following theorem:

\begin{theorem} \label{theo.kag}
Let $T=T(3L,3L)$ be a triangulation of the torus with $L\in\N$. 
Then there are at least two Kempe equivalence classes of edge 
three-colourings of\/ $T$.
In other words, the WSK dynamics for the three-state Potts antiferromagnet 
at zero temperature on the kagom\'e graph $M(T)=T'(3L,3L)$ with $L\in\N$ is 
not ergodic.
\end{theorem}

\proof
The basic strategy is similar to that of the proof of Theorem~3.5
of Ref.~\cite{Mohar_Salas}: We will explicitly construct a four-colouring of 
$T''(3L,3L)=T(6L,6L)$ with the desired properties, and then, we 
apply Lemma~\ref{lemma.kag}. To avoid ambiguities in the computation of the
degree, we orient $T(6L,6L)$ and $\partial \Delta^3$ in such a way that 
the boundary
of all triangular faces are always followed clockwise. The contribution of 
a triangular face $t$ of $T(6L,6L)$ to the degree of a given colouring $f$ 
is $+1$ (resp.\/ $-1$) if
the colouring is $123$ (resp.\/ $132$) if we move clockwise around the boundary
of $t$. In our figures, those faces with orientation preserved (resp.\/ 
reversed) by $f$ are depicted in light (resp.\/ dark) gray.  
We split the proof in two cases, depending on the parity of $L$.

The simpler case is when $L$ is odd, i.e., $3L=6k-3$ for an integer $k\ge 1$.
We only need to prove that there exists a special four-colouring 
$f\in \widetilde{\mathcal{C}}_4(T)$ of the triangulation 
$T=T''(6k-3,6k-3)=T(12k-6,12k-6)$ with $\deg(f)\equiv 6 \pmod{12}$.
This four-colouring is just the  
standard non-singular four-colouring $c_{\hbox{\rm ns}}$. In particular, for
a generic triangulation $T(3L,3M)$, this non-singular four-colouring 
$c_{\hbox{\rm ns}}$ is given by: 
\begin{equation} \fl
c_{\hbox{\rm ns}}(x,y) \;=\; \left\{\begin{array}{ll}
                1 &  \hbox{\rm if $x,y\equiv 1 \bmod{2}$} \\
                4 &  \hbox{\rm if $x  \equiv 1$ and
                              $y  \equiv 0 \bmod{2}$} \\
                2 &  \hbox{\rm if $x  \equiv 0$  and
                              $y  \equiv 1 \bmod{2}$ }\\
                3 &  \hbox{\rm if $x,y\equiv 0 \bmod{2}$}
\end{array}\right. \,, \quad 1\leq x \leq 3L\,, 1\leq y \leq 3M\,.
\label{def_colouring_ns}
\end{equation} 
Its existence for every triangulation $T(12k-6,12k-6)$ with $k\in\N$ is 
given by Proposition~3.2 of Ref.~\cite{Mohar_Salas}.
In addition, $c_{\hbox{\rm ns}}$ for $T=T(12k-6,12k-6)$ has all vertices of 
$V(6k-3,6k-3)$ (and only these) coloured $4$; therefore, it belongs to 
the restricted set $\widetilde{\mathcal{C}}_4(T)$. Finally, its degree 
is given by $\deg(c_{\hbox{\rm ns}})=2(6k-3)^2 \equiv 6 \pmod{12}$.

\bigskip

\noindent
{\bf Remark}. The proper 3--colouring on $T'(6k-3,6k-3)$ associated to
the special 4--coloring $c_{\hbox{\rm ns}}$ on $T(12k-6,12k-6)$  
corresponds to the so-called $Q=0$ state in the physics literature 
\cite{Chalker}.

\bigskip

%
%
\begin{figure}[htb]
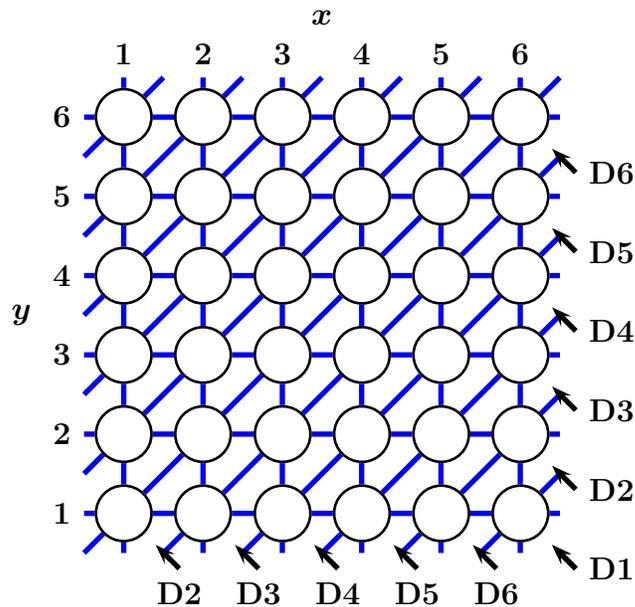

\centering
%
%
\psset{xunit=30pt}
\psset{yunit=30pt}
\pspicture(-1,-1)(6,6.5)
\psline[linewidth=2pt,linecolor=blue](-0.5,0)(5.5,0)
\psline[linewidth=2pt,linecolor=blue](-0.5,1)(5.5,1)
\psline[linewidth=2pt,linecolor=blue](-0.5,2)(5.5,2)
\psline[linewidth=2pt,linecolor=blue](-0.5,3)(5.5,3)
\psline[linewidth=2pt,linecolor=blue](-0.5,4)(5.5,4)
\psline[linewidth=2pt,linecolor=blue](-0.5,5)(5.5,5)
\psline[linewidth=2pt,linecolor=blue](0,-0.5)(0,5.5)
\psline[linewidth=2pt,linecolor=blue](1,-0.5)(1,5.5)
\psline[linewidth=2pt,linecolor=blue](2,-0.5)(2,5.5)
\psline[linewidth=2pt,linecolor=blue](3,-0.5)(3,5.5)
\psline[linewidth=2pt,linecolor=blue](4,-0.5)(4,5.5)
\psline[linewidth=2pt,linecolor=blue](5,-0.5)(5,5.5)
\psline[linewidth=2pt,linecolor=blue](-0.5,-0.5)(5.5,5.5)
\psline[linewidth=2pt,linecolor=blue](0.5,-0.5)(5.5,4.5)
\psline[linewidth=2pt,linecolor=blue](1.5,-0.5)(5.5,3.5)
\psline[linewidth=2pt,linecolor=blue](2.5,-0.5)(5.5,2.5)
\psline[linewidth=2pt,linecolor=blue](3.5,-0.5)(5.5,1.5)
\psline[linewidth=2pt,linecolor=blue](4.5,-0.5)(5.5,0.5)
\psline[linewidth=2pt,linecolor=blue](-0.5,0.5)(4.5,5.5)
\psline[linewidth=2pt,linecolor=blue](-0.5,1.5)(3.5,5.5)
\psline[linewidth=2pt,linecolor=blue](-0.5,2.5)(2.5,5.5)
\psline[linewidth=2pt,linecolor=blue](-0.5,3.5)(1.5,5.5)
\psline[linewidth=2pt,linecolor=blue](-0.5,4.5)(0.5,5.5)
\multirput{0}(0,0)(0,1){6}{%
  \multirput{0}(0,0)(1,0){6}{%
     \pscircle*[linecolor=white]{11pt}
     \pscircle[linewidth=1pt,linecolor=black] {11pt}
   }
}
%
%
\multirput{0}(5.5,-0.5)(0,1){6}{%
     \psline[linewidth=2pt,linecolor=black]{->}(0.2,-0.2)(-0.1,0.1)
}
\uput[0](5.7,-0.7){\bf D1}
\uput[0](5.7, 0.3){\bf D2}
\uput[0](5.7, 1.3){\bf D3}
\uput[0](5.7, 2.3){\bf D4}
\uput[0](5.7, 3.3){\bf D5}
\uput[0](5.7, 4.3){\bf D6}
\multirput{0}(5.5,-0.5)(-1,0){6}{%
     \psline[linewidth=2pt,linecolor=black]{->}(0.2,-0.2)(-0.1,0.1)
}
\uput[270](4.7,-0.7){\bf D6}
\uput[270](3.7,-0.7){\bf D5}
\uput[270](2.7,-0.7){\bf D4}
\uput[270](1.7,-0.7){\bf D3}
\uput[270](0.7,-0.7){\bf D2}
\uput[90](2.5,6){$\bm{x}$}
\uput[90](0,5.5){\bf 1}
\uput[90](1,5.5){\bf 2}
\uput[90](2,5.5){\bf 3}
\uput[90](3,5.5){\bf 4}
\uput[90](4,5.5){\bf 5}
\uput[90](5,5.5){\bf 6}
\uput[180](-0.5,0){\bf 1}
\uput[180](-0.5,1){\bf 2}
\uput[180](-0.5,2){\bf 3}
\uput[180](-0.5,3){\bf 4}
\uput[180](-0.5,4){\bf 5}
\uput[180](-0.5,5){\bf 6}
\uput[180](-1,2.5){$\bm{y}$}
\endpspicture
\caption{\label{figure_tri_notation}
Notation used in the proof of Theorem~\protect\ref{theo.kag}.
Given a triangulation $T(M,M)$ (here we depict the case $M=6$),
we label each vertex using Cartesian coordinates $(x,y)$, $1\leq x,y\leq M$.
The arrows (pointing north-west) show the counter-diagonals D$j$ 
with $j=1,\ldots,M$.
}
\end{figure}
%
%

\proofofcase{2}{$L=2k$}
The above proof does not work for the triangulations $T(6k,6k)$, as 
the non-singular four-colouring of $T''(6k,6k)=T(12k,12k) := T$ has degree
$\equiv 0 \pmod{12}$. We will describe the required four-colouring of $T$
by a construction made in four steps. 
The idea is to build the target four-colouring 
by using counter-diagonals of the triangular lattice: these 
counter-diagonals are orthogonal to the inclined edges of the triangulation
when embedded on a square grid, and will be denoted D$j$, $j=1,2,\dots,12k$. In 
Figure~\ref{figure_tri_notation} we show the triangulation $T(6,6)$, along
with its six counter-diagonals D$j$. As we have embedded the triangulation
into a square grid, we will use Cartesian coordinates 
$(x,y)$, $1\le x,y\le 3L$, for labelling the vertices. 

Let us consider the triangulation $T=T(12k,12k)$ with $k\in\N$. (We will
illustrate the main steps with the case $k=1$). Our construction consists of
four steps: 

\medskip
\noindent
{\bf Step 0.}
To simplify the notation, let us first colour $4$ those vertices in the vertex 
set $V'=V(T(6k,6k))\subset V(T)$. In our standard representation of the 
triangulation $T$ as a square grid with diagonal edges, 
we see that the vertex located at $(x,y)$, $1\leq x,y\leq 12k$, belongs to 
$V'$ if and only if $x\equiv 1 \pmod{2}$ and $y\equiv 0 \pmod{2}$. 

\medskip
\noindent
{\bf Step 1.}
On the counter-diagonal D1 we colour $2$ the $6k$ vertices not already 
coloured $4$.  
On D2, we colour $1$ the vertices with $x$-coordinates either equal to $x=1$
or $6k+1\leq x \leq 12k$. The other $6k-1$ vertices on D2 are coloured $3$.
On D$(12k)$, we colour all vertices $1$ or $3$ in such a way that the resulting 
colouring is proper (for each vertex there is a unique choice).

\begin{figure}[hbt]
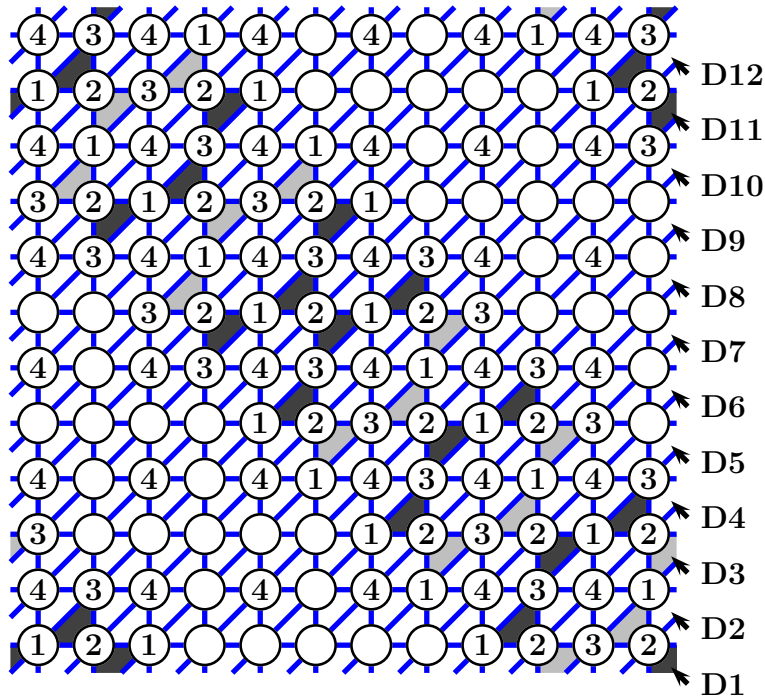

\centering
%
%
\psset{xunit=21pt}
\psset{yunit=21pt}
\psset{labelsep=5pt}
\pspicture(-0.5,-0.9)(12.6,11.5)
\psline*[linewidth=2pt,linecolor=darkgray](0,0)(1,0)(1,1)(0,0)
\psline*[linewidth=2pt,linecolor=lightgray](0,8)(1,8)(1,9)(0,8)
\psline*[linewidth=2pt,linecolor=darkgray](0,10)(1,10)(1,11)(0,10)
\psline*[linewidth=2pt,linecolor=darkgray](1,7)(1,8)(2,8)(1,7)
\psline*[linewidth=2pt,linecolor=lightgray](1,9)(1,10)(2,10)(1,9)
\psline*[linewidth=2pt,linecolor=darkgray](1,11)(1,11.5)(1.5,11.5)(1,11)
\psline*[linewidth=2pt,linecolor=darkgray](1,0)(1,-0.5)(1.5,-0.5)(2,0)(1,0)
\psline*[linewidth=2pt,linecolor=lightgray](2,6)(3,6)(3,7)(2,6)
\psline*[linewidth=2pt,linecolor=darkgray](2,8)(3,8)(3,9)(2,8)
\psline*[linewidth=2pt,linecolor=lightgray](2,10)(3,10)(3,11)(2,10)
\psline*[linewidth=2pt,linecolor=darkgray](3,5)(3,6)(4,6)(3,5)
\psline*[linewidth=2pt,linecolor=lightgray](3,7)(3,8)(4,8)(3,7)
\psline*[linewidth=2pt,linecolor=darkgray](3,9)(3,10)(4,10)(3,9)
\psline*[linewidth=2pt,linecolor=darkgray](4,4)(5,4)(5,5)(4,4)
\psline*[linewidth=2pt,linecolor=darkgray](4,6)(5,6)(5,7)(4,6)
\psline*[linewidth=2pt,linecolor=lightgray](4,8)(5,8)(5,9)(4,8)
\psline*[linewidth=2pt,linecolor=lightgray](5,3)(5,4)(6,4)(5,3)
\psline*[linewidth=2pt,linecolor=darkgray](5,5)(5,6)(6,6)(5,5)
\psline*[linewidth=2pt,linecolor=darkgray](5,7)(5,8)(6,8)(5,7)
\psline*[linewidth=2pt,linecolor=darkgray](6,2)(7,2)(7,3)(6,2)
\psline*[linewidth=2pt,linecolor=lightgray](6,4)(7,4)(7,5)(6,4)
\psline*[linewidth=2pt,linecolor=darkgray](6,6)(7,6)(7,7)(6,6)
\psline*[linewidth=2pt,linecolor=lightgray](7,1)(7,2)(8,2)(7,1)
\psline*[linewidth=2pt,linecolor=darkgray](7,3)(7,4)(8,4)(7,3)
\psline*[linewidth=2pt,linecolor=lightgray](7,5)(7,6)(8,6)(7,5)
\psline*[linewidth=2pt,linecolor=darkgray](8,0)(9,0)(9,1)(8,0)
\psline*[linewidth=2pt,linecolor=lightgray](8,2)(9,2)(9,3)(8,2)
\psline*[linewidth=2pt,linecolor=darkgray](8,4)(9,4)(9,5)(8,4)
\psline*[linewidth=2pt,linecolor=darkgray](9,1)(9,2)(10,2)(9,1)
\psline*[linewidth=2pt,linecolor=lightgray](9,3)(9,4)(10,4)(9,3)
\psline*[linewidth=2pt,linecolor=lightgray](9,11)(9,11.5)(9.5,11.5)(9,11)
\psline*[linewidth=2pt,linecolor=lightgray](9,0)(9,-0.5)(9.5,-0.5)(10,0)(9,0)
\psline*[linewidth=2pt,linecolor=lightgray](10,0)(11,0)(11,1)(10,0)
\psline*[linewidth=2pt,linecolor=darkgray](10,2)(11,2)(11,3)(10,2)
\psline*[linewidth=2pt,linecolor=darkgray](10,10)(11,10)(11,11)(10,10)
\psline*[linewidth=2pt,linecolor=lightgray](11,1)(11,2)(11.5,2)(11.5,1.5)(11,1)
\psline*[linewidth=2pt,linecolor=lightgray](0,2)(-0.5,2)(-0.5,1.5)(0,2)
\psline*[linewidth=2pt,linecolor=darkgray](11,9)(11,10)(11.5,10)(11.5,9.5)(11,9)
\psline*[linewidth=2pt,linecolor=darkgray](0,10)(-0.5,10)(-0.5,9.5)(0,10)
\psline*[linewidth=2pt,linecolor=darkgray](11,11)(11,11.5)(11.5,11.5)(11,11)
\psline*[linewidth=2pt,linecolor=darkgray](11,0)(11.5,0)(11.5,-0.5)(11,-0.5)(11,0)
\psline*[linewidth=2pt,linecolor=darkgray](0,0)(-0.5,0)(-0.5,-0.5)(0,0)
\psline[linewidth=2pt,linecolor=blue](-0.5,0)(11.5,0)
\psline[linewidth=2pt,linecolor=blue](-0.5,1)(11.5,1)
\psline[linewidth=2pt,linecolor=blue](-0.5,2)(11.5,2)
\psline[linewidth=2pt,linecolor=blue](-0.5,3)(11.5,3)
\psline[linewidth=2pt,linecolor=blue](-0.5,4)(11.5,4)
\psline[linewidth=2pt,linecolor=blue](-0.5,5)(11.5,5)
\psline[linewidth=2pt,linecolor=blue](-0.5,6)(11.5,6)
\psline[linewidth=2pt,linecolor=blue](-0.5,7)(11.5,7)
\psline[linewidth=2pt,linecolor=blue](-0.5,8)(11.5,8)
\psline[linewidth=2pt,linecolor=blue](-0.5,9)(11.5,9)
\psline[linewidth=2pt,linecolor=blue](-0.5,10)(11.5,10)
\psline[linewidth=2pt,linecolor=blue](-0.5,11)(11.5,11)
\psline[linewidth=2pt,linecolor=blue](0,-0.5)(0,11.5)
\psline[linewidth=2pt,linecolor=blue](1,-0.5)(1,11.5)
\psline[linewidth=2pt,linecolor=blue](2,-0.5)(2,11.5)
\psline[linewidth=2pt,linecolor=blue](3,-0.5)(3,11.5)
\psline[linewidth=2pt,linecolor=blue](4,-0.5)(4,11.5)
\psline[linewidth=2pt,linecolor=blue](5,-0.5)(5,11.5)
\psline[linewidth=2pt,linecolor=blue](6,-0.5)(6,11.5)
\psline[linewidth=2pt,linecolor=blue](7,-0.5)(7,11.5)
\psline[linewidth=2pt,linecolor=blue](8,-0.5)(8,11.5)
\psline[linewidth=2pt,linecolor=blue](9,-0.5)(9,11.5)
\psline[linewidth=2pt,linecolor=blue](10,-0.5)(10,11.5)
\psline[linewidth=2pt,linecolor=blue](11,-0.5)(11,11.5)
\psline[linewidth=2pt,linecolor=blue](-0.5,-0.5)(11.5,11.5)
\psline[linewidth=2pt,linecolor=blue](0.5,-0.5)(11.5,10.5)
\psline[linewidth=2pt,linecolor=blue](1.5,-0.5)(11.5,9.5)
\psline[linewidth=2pt,linecolor=blue](2.5,-0.5)(11.5,8.5)
\psline[linewidth=2pt,linecolor=blue](3.5,-0.5)(11.5,7.5)
\psline[linewidth=2pt,linecolor=blue](4.5,-0.5)(11.5,6.5)
\psline[linewidth=2pt,linecolor=blue](5.5,-0.5)(11.5,5.5)
\psline[linewidth=2pt,linecolor=blue](6.5,-0.5)(11.5,4.5)
\psline[linewidth=2pt,linecolor=blue](7.5,-0.5)(11.5,3.5)
\psline[linewidth=2pt,linecolor=blue](8.5,-0.5)(11.5,2.5)
\psline[linewidth=2pt,linecolor=blue](9.5,-0.5)(11.5,1.5)
\psline[linewidth=2pt,linecolor=blue](10.5,-0.5)(11.5,0.5)
\psline[linewidth=2pt,linecolor=blue](-0.5,0.5)(10.5,11.5)
\psline[linewidth=2pt,linecolor=blue](-0.5,1.5)(9.5,11.5)
\psline[linewidth=2pt,linecolor=blue](-0.5,2.5)(8.5,11.5)
\psline[linewidth=2pt,linecolor=blue](-0.5,3.5)(7.5,11.5)
\psline[linewidth=2pt,linecolor=blue](-0.5,4.5)(6.5,11.5)
\psline[linewidth=2pt,linecolor=blue](-0.5,5.5)(5.5,11.5)
\psline[linewidth=2pt,linecolor=blue](-0.5,6.5)(4.5,11.5)
\psline[linewidth=2pt,linecolor=blue](-0.5,7.5)(3.5,11.5)
\psline[linewidth=2pt,linecolor=blue](-0.5,8.5)(2.5,11.5)
\psline[linewidth=2pt,linecolor=blue](-0.5,9.5)(1.5,11.5)
\psline[linewidth=2pt,linecolor=blue](-0.5,10.5)(0.5,11.5)
\multirput{0}(0,0)(0,1){12}{%
  \multirput{0}(0,0)(1,0){12}{%
     \pscircle*[linecolor=white]{8pt}
     \pscircle[linewidth=1pt,linecolor=black] {8pt}
   }
}
\rput{0}(0,0){{\bf 1}}
\rput{0}(0,1){{\bf 4}}
\rput{0}(0,2){{\bf 3}}
\rput{0}(0,3){{\bf 4}}
\rput{0}(0,4){{\bf }}
\rput{0}(0,5){{\bf 4}}
\rput{0}(0,6){{\bf }}
\rput{0}(0,7){{\bf 4}}
\rput{0}(0,8){{\bf 3}}
\rput{0}(0,9){{\bf 4}}
\rput{0}(0,10){{\bf 1}}
\rput{0}(0,11){{\bf 4}}
\rput{0}(1,0){{\bf 2}}
\rput{0}(1,1){{\bf 3}}
\rput{0}(1,2){{\bf }}
\rput{0}(1,3){{\bf }}
\rput{0}(1,4){{\bf }}
\rput{0}(1,5){{\bf }}
\rput{0}(1,6){{\bf }}
\rput{0}(1,7){{\bf 3}}
\rput{0}(1,8){{\bf 2}}
\rput{0}(1,9){{\bf 1}}
\rput{0}(1,10){{\bf 2}}
\rput{0}(1,11){{\bf 3}}
\rput{0}(2,0){{\bf 1}}
\rput{0}(2,1){{\bf 4}}
\rput{0}(2,2){{\bf }}
\rput{0}(2,3){{\bf 4}}
\rput{0}(2,4){{\bf }}
\rput{0}(2,5){{\bf 4}}
\rput{0}(2,6){{\bf 3}}
\rput{0}(2,7){{\bf 4}}
\rput{0}(2,8){{\bf 1}}
\rput{0}(2,9){{\bf 4}}
\rput{0}(2,10){{\bf 3}}
\rput{0}(2,11){{\bf 4}}
\rput{0}(3,0){{\bf }}
\rput{0}(3,1){{\bf }}
\rput{0}(3,2){{\bf }}
\rput{0}(3,3){{\bf }}
\rput{0}(3,4){{\bf }}
\rput{0}(3,5){{\bf 3}}
\rput{0}(3,6){{\bf 2}}
\rput{0}(3,7){{\bf 1}}
\rput{0}(3,8){{\bf 2}}
\rput{0}(3,9){{\bf 3}}
\rput{0}(3,10){{\bf 2}}
\rput{0}(3,11){{\bf 1}}
\rput{0}(4,0){{\bf }}
\rput{0}(4,1){{\bf 4}}
\rput{0}(4,2){{\bf }}
\rput{0}(4,3){{\bf 4}}
\rput{0}(4,4){{\bf 1}}
\rput{0}(4,5){{\bf 4}}
\rput{0}(4,6){{\bf 1}}
\rput{0}(4,7){{\bf 4}}
\rput{0}(4,8){{\bf 3}}
\rput{0}(4,9){{\bf 4}}
\rput{0}(4,10){{\bf 1}}
\rput{0}(4,11){{\bf 4}}
\rput{0}(5,0){{\bf }}
\rput{0}(5,1){{\bf }}
\rput{0}(5,2){{\bf }}
\rput{0}(5,3){{\bf 1}}
\rput{0}(5,4){{\bf 2}}
\rput{0}(5,5){{\bf 3}}
\rput{0}(5,6){{\bf 2}}
\rput{0}(5,7){{\bf 3}}
\rput{0}(5,8){{\bf 2}}
\rput{0}(5,9){{\bf 1}}
\rput{0}(5,10){{\bf }}
\rput{0}(5,11){{\bf }}
\rput{0}(6,0){{\bf }}
\rput{0}(6,1){{\bf 4}}
\rput{0}(6,2){{\bf 1}}
\rput{0}(6,3){{\bf 4}}
\rput{0}(6,4){{\bf 3}}
\rput{0}(6,5){{\bf 4}}
\rput{0}(6,6){{\bf 1}}
\rput{0}(6,7){{\bf 4}}
\rput{0}(6,8){{\bf 1}}
\rput{0}(6,9){{\bf 4}}
\rput{0}(6,10){{\bf }}
\rput{0}(6,11){{\bf 4}}
\rput{0}(7,0){{\bf }}
\rput{0}(7,1){{\bf 1}}
\rput{0}(7,2){{\bf 2}}
\rput{0}(7,3){{\bf 3}}
\rput{0}(7,4){{\bf 2}}
\rput{0}(7,5){{\bf 1}}
\rput{0}(7,6){{\bf 2}}
\rput{0}(7,7){{\bf 3}}
\rput{0}(7,8){{\bf }}
\rput{0}(7,9){{\bf }}
\rput{0}(7,10){{\bf }}
\rput{0}(7,11){{\bf }}
\rput{0}(8,0){{\bf 1}}
\rput{0}(8,1){{\bf 4}}
\rput{0}(8,2){{\bf 3}}
\rput{0}(8,3){{\bf 4}}
\rput{0}(8,4){{\bf 1}}
\rput{0}(8,5){{\bf 4}}
\rput{0}(8,6){{\bf 3}}
\rput{0}(8,7){{\bf 4}}
\rput{0}(8,8){{\bf }}
\rput{0}(8,9){{\bf 4}}
\rput{0}(8,10){{\bf }}
\rput{0}(8,11){{\bf 4}}
\rput{0}(9,0){{\bf 2}}
\rput{0}(9,1){{\bf 3}}
\rput{0}(9,2){{\bf 2}}
\rput{0}(9,3){{\bf 1}}
\rput{0}(9,4){{\bf 2}}
\rput{0}(9,5){{\bf 3}}
\rput{0}(9,6){{\bf }}
\rput{0}(9,7){{\bf }}
\rput{0}(9,8){{\bf }}
\rput{0}(9,9){{\bf }}
\rput{0}(9,10){{\bf }}
\rput{0}(9,11){{\bf 1}}
\rput{0}(10,0){{\bf 3}}
\rput{0}(10,1){{\bf 4}}
\rput{0}(10,2){{\bf 1}}
\rput{0}(10,3){{\bf 4}}
\rput{0}(10,4){{\bf 3}}
\rput{0}(10,5){{\bf 4}}
\rput{0}(10,6){{\bf }}
\rput{0}(10,7){{\bf 4}}
\rput{0}(10,8){{\bf }}
\rput{0}(10,9){{\bf 4}}
\rput{0}(10,10){{\bf 1}}
\rput{0}(10,11){{\bf 4}}
\rput{0}(11,0){{\bf 2}}
\rput{0}(11,1){{\bf 1}}
\rput{0}(11,2){{\bf 2}}
\rput{0}(11,3){{\bf 3}}
\rput{0}(11,4){{\bf }}
\rput{0}(11,5){{\bf }}
\rput{0}(11,6){{\bf }}
\rput{0}(11,7){{\bf }}
\rput{0}(11,8){{\bf }}
\rput{0}(11,9){{\bf 3}}
\rput{0}(11,10){{\bf 2}}
\rput{0}(11,11){{\bf 3}}
\multirput{0}(11.5,-0.5)(0,1){12}{%
   \psline[linewidth=2pt,linecolor=black]{->}(0.2,-0.2)(-0.1,0.1)
}
\uput[0](11.7,-0.7){\bf D1}
\uput[0](11.7,0.3){\bf D2}
\uput[0](11.7,1.3){\bf D3}
\uput[0](11.7,2.3){\bf D4}
\uput[0](11.7,3.3){\bf D5}
\uput[0](11.7,4.3){\bf D6}
\uput[0](11.7,5.3){\bf D7}
\uput[0](11.7,6.3){\bf D8}
\uput[0](11.7,7.3){\bf D9}
\uput[0](11.7,8.3){\bf D10}
\uput[0](11.7,9.3){\bf D11}
\uput[0](11.7,10.3){\bf D12}
\endpspicture
\caption{ \label{prop.kag.12k.fig1}
The four-colouring of $T(12,12)$ after the first step of the algorithm given
by the proof of Case~2 of Theorem~\protect\ref{theo.kag}. All vertices 
belonging to the set $V'$ are already coloured~$4$.
}
\end{figure}

We colour all vertices on D3 and D$(12k-1)$ using colour $2$, except those
vertices belonging to $V'$. Finally, we colour all vertices on
D4 and D$(12k-2)$ using colours $1$ and $3$ (again, for each vertex the
choice is unique).
The resulting colouring is depicted on Figure~\ref{prop.kag.12k.fig1}. 
Currently, the partial degree of $f$ is $\deg f|_R = 6$, where we define the
{\em partial degree\/} as the contribution of all triangles already coloured 
$123$ (contribution $+1$) or $132$ (contribution $-1$) towards the degree
of the targeted colouring $f$. 

%
%
%
%

\medskip
\noindent
{\bf Step 2.}
There are $12k-7$ counter-diagonals to be coloured, and in this step
we will sequentially colour all of them but five. 
(Observe that there is nothing to do if $k=1$.)
This will be done by performing the following procedure: Suppose that we
have already coloured counter-diagonals D$j$ and D$(12k-j+2)$ ($j\geq 4$) 
using colours $1$ and $3$. We first colour D$(j+1)$ and D$(12k-j+1)$ using 
colour $2$ for all vertices not belonging to $V'$. 
We then colour D$(j+2)$ and D$(12k-j)$ using colours $1$ and $3$. In these
cases, for each vertex we have only one choice. This step is repeated 
$3(k-1)$ times: We add $12(k-1)$ counter-diagonals, and there are only
five counter-diagonals not coloured yet. Indeed, the last coloured 
counter-diagonals use colours $1$ and $3$, as at the end of Step~1. 

Each of these $3(k-1)$ steps adds a $4$ to the degree of the colouring.
Namely, all new triangles coloured 123 or 132 are located along the 
counter-diagonals D$(j+1)$ and D$(12k-j-1)$. Triangles coloured 123 and 132
come in pairs, annihilating each other's contribution, except at the vertex
coloured 2 where colour 3 is changed to 1 on the next counter-diagonal. 
There we get two triangles, each contributing $+1$, on each of the
counter-diagonals D$(j+1)$ and D$(12k-j-1)$.
Thus, the partial degree of the colouring is  $\deg f|_R = 6+12(k-1)$.  

\medskip
\noindent
{\bf Step 3.}
The last coloured counter-diagonals are D$(6k-2)$ and D$(6k+4)$. 

On D$(6k-1)$, there is a single vertex not in $V'$ whose
colour can only be $2$ since it has neighbours of colours 1, 3, and 4.
This vertex is located at 
$x=9k-1$ (resp.\  $x=3k$) if $k$ is odd (resp. even). We then colour
the vertex $v_0=(x_0,y_0)$ on D$(6k-1)$ located at $x_0=3k-1$  
(resp.\ $x_0=9k$) using colour $1$ (resp.\  $3$) if $k$ is odd (resp.\  even). 
The remaining vertices on D$(6k-1)$ are coloured $2$.  

On $D(6k)$ we find that there are two vertices that are neighbours of 
$v_0$ and should be coloured $2$. These vertices are located at
$(x_0,y_0+1)$ and $(x_0+1,y_0)$ (valid for every $k$).
The other vertices on $D(6k)$ are coloured $1$ or $3$; the choice is
unique for each vertex. 
As shown in Figure~\ref{prop.kag.12k.fig2}, the contribution to the degree of 
these new triangular faces is $2$; thus, the partial degree of $f$ is  
$\deg f|_R = 8 +12 (k-1)$.

%
%
\begin{figure}[hbt]
\centering
%
%
\psset{xunit=21pt}
\psset{yunit=21pt}
\psset{labelsep=5pt}
\pspicture(-0.5,-0.9)(12.6,11.5)
\psline*[linewidth=2pt,linecolor=darkgray](0,0)(1,0)(1,1)(0,0)
\psline*[linewidth=2pt,linecolor=darkgray](0,2)(1,2)(1,3)(0,2)
\psline*[linewidth=2pt,linecolor=lightgray](0,8)(1,8)(1,9)(0,8)
\psline*[linewidth=2pt,linecolor=darkgray](0,10)(1,10)(1,11)(0,10)
\psline*[linewidth=2pt,linecolor=lightgray](1,1)(1,2)(2,2)(1,1)
\psline*[linewidth=2pt,linecolor=darkgray](1,7)(1,8)(2,8)(1,7)
\psline*[linewidth=2pt,linecolor=lightgray](1,9)(1,10)(2,10)(1,9)
\psline*[linewidth=2pt,linecolor=darkgray](1,11)(1,11.5)(1.5,11.5)(1,11)
\psline*[linewidth=2pt,linecolor=darkgray](1,0)(1,-0.5)(1.5,-0.5)(2,0)(1,0)
\psline*[linewidth=2pt,linecolor=darkgray](2,0)(3,0)(3,1)(2,0)
\psline*[linewidth=2pt,linecolor=lightgray](2,6)(3,6)(3,7)(2,6)
\psline*[linewidth=2pt,linecolor=darkgray](2,8)(3,8)(3,9)(2,8)
\psline*[linewidth=2pt,linecolor=lightgray](2,10)(3,10)(3,11)(2,10)
\psline*[linewidth=2pt,linecolor=darkgray](3,5)(3,6)(4,6)(3,5)
\psline*[linewidth=2pt,linecolor=lightgray](3,7)(3,8)(4,8)(3,7)
\psline*[linewidth=2pt,linecolor=darkgray](3,9)(3,10)(4,10)(3,9)
\psline*[linewidth=2pt,linecolor=lightgray](3,11)(3,11.5)(3.5,11.5)(3,11)
\psline*[linewidth=2pt,linecolor=lightgray](3,0)(3,-0.5)(3.5,-0.5)(4,0)(3,0)
\psline*[linewidth=2pt,linecolor=darkgray](4,4)(5,4)(5,5)(4,4)
\psline*[linewidth=2pt,linecolor=darkgray](4,6)(5,6)(5,7)(4,6)
\psline*[linewidth=2pt,linecolor=lightgray](4,8)(5,8)(5,9)(4,8)
\psline*[linewidth=2pt,linecolor=darkgray](4,10)(5,10)(5,11)(4,10)
\psline*[linewidth=2pt,linecolor=lightgray](5,3)(5,4)(6,4)(5,3)
\psline*[linewidth=2pt,linecolor=darkgray](5,5)(5,6)(6,6)(5,5)
\psline*[linewidth=2pt,linecolor=darkgray](5,7)(5,8)(6,8)(5,7)
\psline*[linewidth=2pt,linecolor=lightgray](5,9)(5,10)(6,10)(5,9)
\psline*[linewidth=2pt,linecolor=darkgray](6,2)(7,2)(7,3)(6,2)
\psline*[linewidth=2pt,linecolor=lightgray](6,4)(7,4)(7,5)(6,4)
\psline*[linewidth=2pt,linecolor=darkgray](6,6)(7,6)(7,7)(6,6)
\psline*[linewidth=2pt,linecolor=darkgray](6,8)(7,8)(7,9)(6,8)
\psline*[linewidth=2pt,linecolor=lightgray](7,1)(7,2)(8,2)(7,1)
\psline*[linewidth=2pt,linecolor=darkgray](7,3)(7,4)(8,4)(7,3)
\psline*[linewidth=2pt,linecolor=lightgray](7,5)(7,6)(8,6)(7,5)
\psline*[linewidth=2pt,linecolor=darkgray](7,7)(7,8)(8,8)(7,7)
\psline*[linewidth=2pt,linecolor=darkgray](8,0)(9,0)(9,1)(8,0)
\psline*[linewidth=2pt,linecolor=lightgray](8,2)(9,2)(9,3)(8,2)
\psline*[linewidth=2pt,linecolor=darkgray](8,4)(9,4)(9,5)(8,4)
\psline*[linewidth=2pt,linecolor=lightgray](8,6)(9,6)(9,7)(8,6)
\psline*[linewidth=2pt,linecolor=darkgray](9,1)(9,2)(10,2)(9,1)
\psline*[linewidth=2pt,linecolor=lightgray](9,3)(9,4)(10,4)(9,3)
\psline*[linewidth=2pt,linecolor=darkgray](9,5)(9,6)(10,6)(9,5)
\psline*[linewidth=2pt,linecolor=lightgray](9,11)(9,11.5)(9.5,11.5)(9,11)
\psline*[linewidth=2pt,linecolor=lightgray](9,0)(9,-0.5)(9.5,-0.5)(10,0)(9,0)
\psline*[linewidth=2pt,linecolor=lightgray](10,0)(11,0)(11,1)(10,0)
\psline*[linewidth=2pt,linecolor=darkgray](10,2)(11,2)(11,3)(10,2)
\psline*[linewidth=2pt,linecolor=lightgray](10,4)(11,4)(11,5)(10,4)
\psline*[linewidth=2pt,linecolor=darkgray](10,10)(11,10)(11,11)(10,10)
\psline*[linewidth=2pt,linecolor=lightgray](11,1)(11,2)(11.5,2)(11.5,1.5)(11,1)
\psline*[linewidth=2pt,linecolor=lightgray](0,2)(-0.5,2)(-0.5,1.5)(0,2)
\psline*[linewidth=2pt,linecolor=darkgray](11,3)(11,4)(11.5,4)(11.5,3.5)(11,3)
\psline*[linewidth=2pt,linecolor=darkgray](0,4)(-0.5,4)(-0.5,3.5)(0,4)
\psline*[linewidth=2pt,linecolor=darkgray](11,9)(11,10)(11.5,10)(11.5,9.5)(11,9)
\psline*[linewidth=2pt,linecolor=darkgray](0,10)(-0.5,10)(-0.5,9.5)(0,10)
\psline*[linewidth=2pt,linecolor=darkgray](11,11)(11,11.5)(11.5,11.5)(11,11)
\psline*[linewidth=2pt,linecolor=darkgray](11,0)(11.5,0)(11.5,-0.5)(11,-0.5)(11,0)
\psline*[linewidth=2pt,linecolor=darkgray](0,0)(-0.5,0)(-0.5,-0.5)(0,0)
\psline[linewidth=2pt,linecolor=blue](-0.5,0)(11.5,0)
\psline[linewidth=2pt,linecolor=blue](-0.5,1)(11.5,1)
\psline[linewidth=2pt,linecolor=blue](-0.5,2)(11.5,2)
\psline[linewidth=2pt,linecolor=blue](-0.5,3)(11.5,3)
\psline[linewidth=2pt,linecolor=blue](-0.5,4)(11.5,4)
\psline[linewidth=2pt,linecolor=blue](-0.5,5)(11.5,5)
\psline[linewidth=2pt,linecolor=blue](-0.5,6)(11.5,6)
\psline[linewidth=2pt,linecolor=blue](-0.5,7)(11.5,7)
\psline[linewidth=2pt,linecolor=blue](-0.5,8)(11.5,8)
\psline[linewidth=2pt,linecolor=blue](-0.5,9)(11.5,9)
\psline[linewidth=2pt,linecolor=blue](-0.5,10)(11.5,10)
\psline[linewidth=2pt,linecolor=blue](-0.5,11)(11.5,11)
\psline[linewidth=2pt,linecolor=blue](0,-0.5)(0,11.5)
\psline[linewidth=2pt,linecolor=blue](1,-0.5)(1,11.5)
\psline[linewidth=2pt,linecolor=blue](2,-0.5)(2,11.5)
\psline[linewidth=2pt,linecolor=blue](3,-0.5)(3,11.5)
\psline[linewidth=2pt,linecolor=blue](4,-0.5)(4,11.5)
\psline[linewidth=2pt,linecolor=blue](5,-0.5)(5,11.5)
\psline[linewidth=2pt,linecolor=blue](6,-0.5)(6,11.5)
\psline[linewidth=2pt,linecolor=blue](7,-0.5)(7,11.5)
\psline[linewidth=2pt,linecolor=blue](8,-0.5)(8,11.5)
\psline[linewidth=2pt,linecolor=blue](9,-0.5)(9,11.5)
\psline[linewidth=2pt,linecolor=blue](10,-0.5)(10,11.5)
\psline[linewidth=2pt,linecolor=blue](11,-0.5)(11,11.5)
\psline[linewidth=2pt,linecolor=blue](-0.5,-0.5)(11.5,11.5)
\psline[linewidth=2pt,linecolor=blue](0.5,-0.5)(11.5,10.5)
\psline[linewidth=2pt,linecolor=blue](1.5,-0.5)(11.5,9.5)
\psline[linewidth=2pt,linecolor=blue](2.5,-0.5)(11.5,8.5)
\psline[linewidth=2pt,linecolor=blue](3.5,-0.5)(11.5,7.5)
\psline[linewidth=2pt,linecolor=blue](4.5,-0.5)(11.5,6.5)
\psline[linewidth=2pt,linecolor=blue](5.5,-0.5)(11.5,5.5)
\psline[linewidth=2pt,linecolor=blue](6.5,-0.5)(11.5,4.5)
\psline[linewidth=2pt,linecolor=blue](7.5,-0.5)(11.5,3.5)
\psline[linewidth=2pt,linecolor=blue](8.5,-0.5)(11.5,2.5)
\psline[linewidth=2pt,linecolor=blue](9.5,-0.5)(11.5,1.5)
\psline[linewidth=2pt,linecolor=blue](10.5,-0.5)(11.5,0.5)
\psline[linewidth=2pt,linecolor=blue](-0.5,0.5)(10.5,11.5)
\psline[linewidth=2pt,linecolor=blue](-0.5,1.5)(9.5,11.5)
\psline[linewidth=2pt,linecolor=blue](-0.5,2.5)(8.5,11.5)
\psline[linewidth=2pt,linecolor=blue](-0.5,3.5)(7.5,11.5)
\psline[linewidth=2pt,linecolor=blue](-0.5,4.5)(6.5,11.5)
\psline[linewidth=2pt,linecolor=blue](-0.5,5.5)(5.5,11.5)
\psline[linewidth=2pt,linecolor=blue](-0.5,6.5)(4.5,11.5)
\psline[linewidth=2pt,linecolor=blue](-0.5,7.5)(3.5,11.5)
\psline[linewidth=2pt,linecolor=blue](-0.5,8.5)(2.5,11.5)
\psline[linewidth=2pt,linecolor=blue](-0.5,9.5)(1.5,11.5)
\psline[linewidth=2pt,linecolor=blue](-0.5,10.5)(0.5,11.5)
\multirput{0}(0,0)(0,1){12}{%
  \multirput{0}(0,0)(1,0){12}{%
     \pscircle*[linecolor=white]{8pt}
     \pscircle[linewidth=1pt,linecolor=black]{8pt}
   }
}
\rput{0}(0,0){{\bf 1}}
\rput{0}(0,1){{\bf 4}}
\rput{0}(0,2){{\bf 3}}
\rput{0}(0,3){{\bf 4}}
\rput{0}(0,4){{\bf 1}}
\rput{0}(0,5){{\bf 4}}
\rput{0}(0,6){{\bf }}
\rput{0}(0,7){{\bf 4}}
\rput{0}(0,8){{\bf 3}}
\rput{0}(0,9){{\bf 4}}
\rput{0}(0,10){{\bf 1}}
\rput{0}(0,11){{\bf 4}}
\rput{0}(1,0){{\bf 2}}
\rput{0}(1,1){{\bf 3}}
\rput{0}(1,2){{\bf 1}}
\rput{0}(1,3){{\bf 2}}
\rput{0}(1,4){{\bf }}
\rput{0}(1,5){{\bf }}
\rput{0}(1,6){{\bf }}
\rput{0}(1,7){{\bf 3}}
\rput{0}(1,8){{\bf 2}}
\rput{0}(1,9){{\bf 1}}
\rput{0}(1,10){{\bf 2}}
\rput{0}(1,11){{\bf 3}}
\rput{0}(2,0){{\bf 1}}
\rput{0}(2,1){{\bf 4}}
\rput{0}(2,2){{\bf 2}}
\rput{0}(2,3){{\bf 4}}
\rput{0}(2,4){{\bf }}
\rput{0}(2,5){{\bf 4}}
\rput{0}(2,6){{\bf 3}}
\rput{0}(2,7){{\bf 4}}
\rput{0}(2,8){{\bf 1}}
\rput{0}(2,9){{\bf 4}}
\rput{0}(2,10){{\bf 3}}
\rput{0}(2,11){{\bf 4}}
\rput{0}(3,0){{\bf 2}}
\rput{0}(3,1){{\bf 3}}
\rput{0}(3,2){{\bf }}
\rput{0}(3,3){{\bf }}
\rput{0}(3,4){{\bf }}
\rput{0}(3,5){{\bf 3}}
\rput{0}(3,6){{\bf 2}}
\rput{0}(3,7){{\bf 1}}
\rput{0}(3,8){{\bf 2}}
\rput{0}(3,9){{\bf 3}}
\rput{0}(3,10){{\bf 2}}
\rput{0}(3,11){{\bf 1}}
\rput{0}(4,0){{\bf 3}}
\rput{0}(4,1){{\bf 4}}
\rput{0}(4,2){{\bf }}
\rput{0}(4,3){{\bf 4}}
\rput{0}(4,4){{\bf 1}}
\rput{0}(4,5){{\bf 4}}
\rput{0}(4,6){{\bf 1}}
\rput{0}(4,7){{\bf 4}}
\rput{0}(4,8){{\bf 3}}
\rput{0}(4,9){{\bf 4}}
\rput{0}(4,10){{\bf 1}}
\rput{0}(4,11){{\bf 4}}
\rput{0}(5,0){{\bf }}
\rput{0}(5,1){{\bf }}
\rput{0}(5,2){{\bf }}
\rput{0}(5,3){{\bf 1}}
\rput{0}(5,4){{\bf 2}}
\rput{0}(5,5){{\bf 3}}
\rput{0}(5,6){{\bf 2}}
\rput{0}(5,7){{\bf 3}}
\rput{0}(5,8){{\bf 2}}
\rput{0}(5,9){{\bf 1}}
\rput{0}(5,10){{\bf 2}}
\rput{0}(5,11){{\bf 3}}
\rput{0}(6,0){{\bf }}
\rput{0}(6,1){{\bf 4}}
\rput{0}(6,2){{\bf 1}}
\rput{0}(6,3){{\bf 4}}
\rput{0}(6,4){{\bf 3}}
\rput{0}(6,5){{\bf 4}}
\rput{0}(6,6){{\bf 1}}
\rput{0}(6,7){{\bf 4}}
\rput{0}(6,8){{\bf 1}}
\rput{0}(6,9){{\bf 4}}
\rput{0}(6,10){{\bf 3}}
\rput{0}(6,11){{\bf 4}}
\rput{0}(7,0){{\bf }}
\rput{0}(7,1){{\bf 1}}
\rput{0}(7,2){{\bf 2}}
\rput{0}(7,3){{\bf 3}}
\rput{0}(7,4){{\bf 2}}
\rput{0}(7,5){{\bf 1}}
\rput{0}(7,6){{\bf 2}}
\rput{0}(7,7){{\bf 3}}
\rput{0}(7,8){{\bf 2}}
\rput{0}(7,9){{\bf 3}}
\rput{0}(7,10){{\bf }}
\rput{0}(7,11){{\bf }}
\rput{0}(8,0){{\bf 1}}
\rput{0}(8,1){{\bf 4}}
\rput{0}(8,2){{\bf 3}}
\rput{0}(8,3){{\bf 4}}
\rput{0}(8,4){{\bf 1}}
\rput{0}(8,5){{\bf 4}}
\rput{0}(8,6){{\bf 3}}
\rput{0}(8,7){{\bf 4}}
\rput{0}(8,8){{\bf 1}}
\rput{0}(8,9){{\bf 4}}
\rput{0}(8,10){{\bf }}
\rput{0}(8,11){{\bf 4}}
\rput{0}(9,0){{\bf 2}}
\rput{0}(9,1){{\bf 3}}
\rput{0}(9,2){{\bf 2}}
\rput{0}(9,3){{\bf 1}}
\rput{0}(9,4){{\bf 2}}
\rput{0}(9,5){{\bf 3}}
\rput{0}(9,6){{\bf 2}}
\rput{0}(9,7){{\bf 1}}
\rput{0}(9,8){{\bf }}
\rput{0}(9,9){{\bf }}
\rput{0}(9,10){{\bf }}
\rput{0}(9,11){{\bf 1}}
\rput{0}(10,0){{\bf 3}}
\rput{0}(10,1){{\bf 4}}
\rput{0}(10,2){{\bf 1}}
\rput{0}(10,3){{\bf 4}}
\rput{0}(10,4){{\bf 3}}
\rput{0}(10,5){{\bf 4}}
\rput{0}(10,6){{\bf 1}}
\rput{0}(10,7){{\bf 4}}
\rput{0}(10,8){{\bf }}
\rput{0}(10,9){{\bf 4}}
\rput{0}(10,10){{\bf 1}}
\rput{0}(10,11){{\bf 4}}
\rput{0}(11,0){{\bf 2}}
\rput{0}(11,1){{\bf 1}}
\rput{0}(11,2){{\bf 2}}
\rput{0}(11,3){{\bf 3}}
\rput{0}(11,4){{\bf 2}}
\rput{0}(11,5){{\bf 1}}
\rput{0}(11,6){{\bf }}
\rput{0}(11,7){{\bf }}
\rput{0}(11,8){{\bf }}
\rput{0}(11,9){{\bf 3}}
\rput{0}(11,10){{\bf 2}}
\rput{0}(11,11){{\bf 3}}
\multirput{0}(11.5,-0.5)(0,1){12}{%
   \psline[linewidth=2pt,linecolor=black]{->}(0.2,-0.2)(-0.1,0.1)
}
\uput[0](11.7,-0.7){\bf D1}
\uput[0](11.7,0.3){\bf D2}
\uput[0](11.7,1.3){\bf D3}
\uput[0](11.7,2.3){\bf D4}
\uput[0](11.7,3.3){\bf D5}
\uput[0](11.7,4.3){\bf D6}
\uput[0](11.7,5.3){\bf D7}
\uput[0](11.7,6.3){\bf D8}
\uput[0](11.7,7.3){\bf D9}
\uput[0](11.7,8.3){\bf D10}
\uput[0](11.7,9.3){\bf D11}
\uput[0](11.7,10.3){\bf D12}
\endpspicture
\caption{ \label{prop.kag.12k.fig2}
The four-colouring of $T(12,12)$ after Step~3 of the algorithm given
by the proof of Case~2 of Theorem~\protect\ref{theo.kag}
}
\end{figure}
%
%

\medskip
\noindent
{\bf Step 4.}
On D$(6k+1)$ the vertices at $(x_0+2,y_0)$ and $(x_0,y_0+2)$ (where the
special vertex $v_0=(x_0,y_0)$ was defined in the previous step) should
be coloured $1$ and $3$, respectively. The other vertices are coloured
$1$ or $3$ (the choice among these two possible colours is again unique 
for each vertex). 

On D$(6k+3)$, we find that the vertex at $x=x_0+2$ should be coloured $2$.
The other vertices on this counter-diagonal and not belonging to
$V'$ are coloured $1$ or $3$; the choice among these two colours is again unique.

%
%
\begin{figure}[htb]
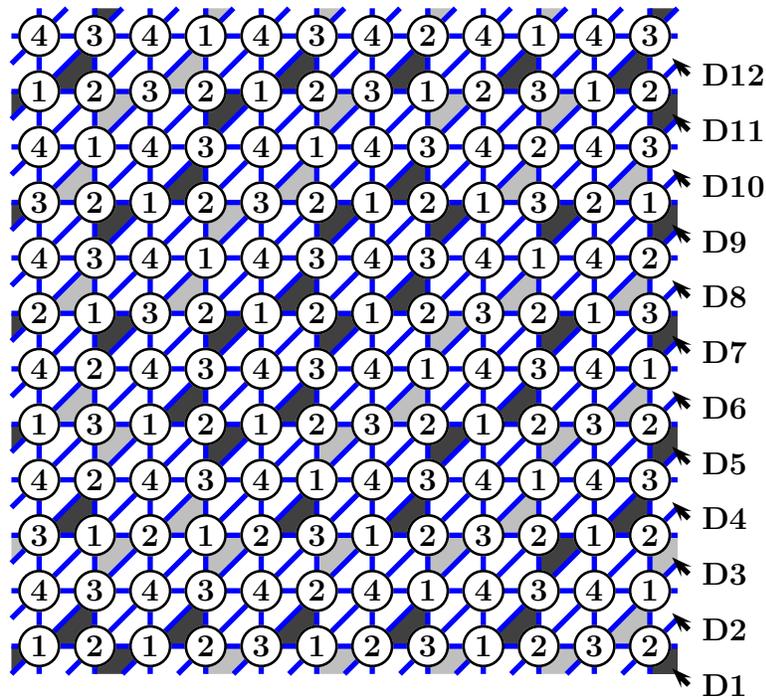

\centering
%
%
\psset{xunit=21pt}
\psset{yunit=21pt}
\psset{labelsep=5pt}
\pspicture(-0.5,-0.9)(12.6,11.5)
\psline*[linewidth=2pt,linecolor=darkgray](0,0)(1,0)(1,1)(0,0)
\psline*[linewidth=2pt,linecolor=darkgray](0,2)(1,2)(1,3)(0,2)
\psline*[linewidth=2pt,linecolor=lightgray](0,4)(1,4)(1,5)(0,4)
\psline*[linewidth=2pt,linecolor=lightgray](0,6)(1,6)(1,7)(0,6)
\psline*[linewidth=2pt,linecolor=lightgray](0,8)(1,8)(1,9)(0,8)
\psline*[linewidth=2pt,linecolor=darkgray](0,10)(1,10)(1,11)(0,10)
\psline*[linewidth=2pt,linecolor=lightgray](1,1)(1,2)(2,2)(1,1)
\psline*[linewidth=2pt,linecolor=lightgray](1,3)(1,4)(2,4)(1,3)
\psline*[linewidth=2pt,linecolor=darkgray](1,5)(1,6)(2,6)(1,5)
\psline*[linewidth=2pt,linecolor=darkgray](1,7)(1,8)(2,8)(1,7)
\psline*[linewidth=2pt,linecolor=lightgray](1,9)(1,10)(2,10)(1,9)
\psline*[linewidth=2pt,linecolor=darkgray](1,11)(1,11.5)(1.5,11.5)(1,11)
\psline*[linewidth=2pt,linecolor=darkgray](1,0)(1,-0.5)(1.5,-0.5)(2,0)(1,0)
\psline*[linewidth=2pt,linecolor=darkgray](2,0)(3,0)(3,1)(2,0)
\psline*[linewidth=2pt,linecolor=lightgray](2,2)(3,2)(3,3)(2,2)
\psline*[linewidth=2pt,linecolor=darkgray](2,4)(3,4)(3,5)(2,4)
\psline*[linewidth=2pt,linecolor=lightgray](2,6)(3,6)(3,7)(2,6)
\psline*[linewidth=2pt,linecolor=darkgray](2,8)(3,8)(3,9)(2,8)
\psline*[linewidth=2pt,linecolor=lightgray](2,10)(3,10)(3,11)(2,10)
\psline*[linewidth=2pt,linecolor=lightgray](3,1)(3,2)(4,2)(3,1)
\psline*[linewidth=2pt,linecolor=darkgray](3,3)(3,4)(4,4)(3,3)
\psline*[linewidth=2pt,linecolor=darkgray](3,5)(3,6)(4,6)(3,5)
\psline*[linewidth=2pt,linecolor=lightgray](3,7)(3,8)(4,8)(3,7)
\psline*[linewidth=2pt,linecolor=darkgray](3,9)(3,10)(4,10)(3,9)
\psline*[linewidth=2pt,linecolor=lightgray](3,11)(3,11.5)(3.5,11.5)(3,11)
\psline*[linewidth=2pt,linecolor=lightgray](3,0)(3,-0.5)(3.5,-0.5)(4,0)(3,0)
\psline*[linewidth=2pt,linecolor=darkgray](4,0)(5,0)(5,1)(4,0)
\psline*[linewidth=2pt,linecolor=darkgray](4,2)(5,2)(5,3)(4,2)
\psline*[linewidth=2pt,linecolor=darkgray](4,4)(5,4)(5,5)(4,4)
\psline*[linewidth=2pt,linecolor=darkgray](4,6)(5,6)(5,7)(4,6)
\psline*[linewidth=2pt,linecolor=lightgray](4,8)(5,8)(5,9)(4,8)
\psline*[linewidth=2pt,linecolor=darkgray](4,10)(5,10)(5,11)(4,10)
\psline*[linewidth=2pt,linecolor=lightgray](5,1)(5,2)(6,2)(5,1)
\psline*[linewidth=2pt,linecolor=lightgray](5,3)(5,4)(6,4)(5,3)
\psline*[linewidth=2pt,linecolor=darkgray](5,5)(5,6)(6,6)(5,5)
\psline*[linewidth=2pt,linecolor=darkgray](5,7)(5,8)(6,8)(5,7)
\psline*[linewidth=2pt,linecolor=lightgray](5,9)(5,10)(6,10)(5,9)
\psline*[linewidth=2pt,linecolor=lightgray](5,11)(5,11.5)(5.5,11.5)(5,11)
\psline*[linewidth=2pt,linecolor=lightgray](5,0)(5,-0.5)(5.5,-0.5)(6,0)(5,0)
\psline*[linewidth=2pt,linecolor=darkgray](6,0)(7,0)(7,1)(6,0)
\psline*[linewidth=2pt,linecolor=darkgray](6,2)(7,2)(7,3)(6,2)
\psline*[linewidth=2pt,linecolor=lightgray](6,4)(7,4)(7,5)(6,4)
\psline*[linewidth=2pt,linecolor=darkgray](6,6)(7,6)(7,7)(6,6)
\psline*[linewidth=2pt,linecolor=darkgray](6,8)(7,8)(7,9)(6,8)
\psline*[linewidth=2pt,linecolor=darkgray](6,10)(7,10)(7,11)(6,10)
\psline*[linewidth=2pt,linecolor=lightgray](7,1)(7,2)(8,2)(7,1)
\psline*[linewidth=2pt,linecolor=darkgray](7,3)(7,4)(8,4)(7,3)
\psline*[linewidth=2pt,linecolor=lightgray](7,5)(7,6)(8,6)(7,5)
\psline*[linewidth=2pt,linecolor=darkgray](7,7)(7,8)(8,8)(7,7)
\psline*[linewidth=2pt,linecolor=lightgray](7,9)(7,10)(8,10)(7,9)
\psline*[linewidth=2pt,linecolor=lightgray](7,11)(7,11.5)(7.5,11.5)(7,11)
\psline*[linewidth=2pt,linecolor=lightgray](7,0)(7,-0.5)(7.5,-0.5)(8,0)(7,0)
\psline*[linewidth=2pt,linecolor=darkgray](8,0)(9,0)(9,1)(8,0)
\psline*[linewidth=2pt,linecolor=lightgray](8,2)(9,2)(9,3)(8,2)
\psline*[linewidth=2pt,linecolor=darkgray](8,4)(9,4)(9,5)(8,4)
\psline*[linewidth=2pt,linecolor=lightgray](8,6)(9,6)(9,7)(8,6)
\psline*[linewidth=2pt,linecolor=lightgray](8,8)(9,8)(9,9)(8,8)
\psline*[linewidth=2pt,linecolor=darkgray](8,10)(9,10)(9,11)(8,10)
\psline*[linewidth=2pt,linecolor=darkgray](9,1)(9,2)(10,2)(9,1)
\psline*[linewidth=2pt,linecolor=lightgray](9,3)(9,4)(10,4)(9,3)
\psline*[linewidth=2pt,linecolor=darkgray](9,5)(9,6)(10,6)(9,5)
\psline*[linewidth=2pt,linecolor=darkgray](9,7)(9,8)(10,8)(9,7)
\psline*[linewidth=2pt,linecolor=lightgray](9,9)(9,10)(10,10)(9,9)
\psline*[linewidth=2pt,linecolor=lightgray](9,11)(9,11.5)(9.5,11.5)(9,11)
\psline*[linewidth=2pt,linecolor=lightgray](9,0)(9,-0.5)(9.5,-0.5)(10,0)(9,0)
\psline*[linewidth=2pt,linecolor=lightgray](10,0)(11,0)(11,1)(10,0)
\psline*[linewidth=2pt,linecolor=darkgray](10,2)(11,2)(11,3)(10,2)
\psline*[linewidth=2pt,linecolor=lightgray](10,4)(11,4)(11,5)(10,4)
\psline*[linewidth=2pt,linecolor=lightgray](10,6)(11,6)(11,7)(10,6)
\psline*[linewidth=2pt,linecolor=lightgray](10,8)(11,8)(11,9)(10,8)
\psline*[linewidth=2pt,linecolor=darkgray](10,10)(11,10)(11,11)(10,10)
\psline*[linewidth=2pt,linecolor=lightgray](11,1)(11,2)(11.5,2)(11.5,1.5)(11,1)
\psline*[linewidth=2pt,linecolor=lightgray](0,2)(-0.5,2)(-0.5,1.5)(0,2)
\psline*[linewidth=2pt,linecolor=darkgray](11,3)(11,4)(11.5,4)(11.5,3.5)(11,3)
\psline*[linewidth=2pt,linecolor=darkgray](0,4)(-0.5,4)(-0.5,3.5)(0,4)
\psline*[linewidth=2pt,linecolor=darkgray](11,5)(11,6)(11.5,6)(11.5,5.5)(11,5)
\psline*[linewidth=2pt,linecolor=darkgray](0,6)(-0.5,6)(-0.5,5.5)(0,6)
\psline*[linewidth=2pt,linecolor=darkgray](11,7)(11,8)(11.5,8)(11.5,7.5)(11,7)
\psline*[linewidth=2pt,linecolor=darkgray](0,8)(-0.5,8)(-0.5,7.5)(0,8)
\psline*[linewidth=2pt,linecolor=darkgray](11,9)(11,10)(11.5,10)(11.5,9.5)(11,9)
\psline*[linewidth=2pt,linecolor=darkgray](0,10)(-0.5,10)(-0.5,9.5)(0,10)
\psline*[linewidth=2pt,linecolor=darkgray](11,11)(11,11.5)(11.5,11.5)(11,11)
\psline*[linewidth=2pt,linecolor=darkgray](11,0)(11.5,0)(11.5,-0.5)(11,-0.5)(11,0)
\psline*[linewidth=2pt,linecolor=darkgray](0,0)(-0.5,0)(-0.5,-0.5)(0,0)
\psline[linewidth=2pt,linecolor=blue](-0.5,0)(11.5,0)
\psline[linewidth=2pt,linecolor=blue](-0.5,1)(11.5,1)
\psline[linewidth=2pt,linecolor=blue](-0.5,2)(11.5,2)
\psline[linewidth=2pt,linecolor=blue](-0.5,3)(11.5,3)
\psline[linewidth=2pt,linecolor=blue](-0.5,4)(11.5,4)
\psline[linewidth=2pt,linecolor=blue](-0.5,5)(11.5,5)
\psline[linewidth=2pt,linecolor=blue](-0.5,6)(11.5,6)
\psline[linewidth=2pt,linecolor=blue](-0.5,7)(11.5,7)
\psline[linewidth=2pt,linecolor=blue](-0.5,8)(11.5,8)
\psline[linewidth=2pt,linecolor=blue](-0.5,9)(11.5,9)
\psline[linewidth=2pt,linecolor=blue](-0.5,10)(11.5,10)
\psline[linewidth=2pt,linecolor=blue](-0.5,11)(11.5,11)
\psline[linewidth=2pt,linecolor=blue](0,-0.5)(0,11.5)
\psline[linewidth=2pt,linecolor=blue](1,-0.5)(1,11.5)
\psline[linewidth=2pt,linecolor=blue](2,-0.5)(2,11.5)
\psline[linewidth=2pt,linecolor=blue](3,-0.5)(3,11.5)
\psline[linewidth=2pt,linecolor=blue](4,-0.5)(4,11.5)
\psline[linewidth=2pt,linecolor=blue](5,-0.5)(5,11.5)
\psline[linewidth=2pt,linecolor=blue](6,-0.5)(6,11.5)
\psline[linewidth=2pt,linecolor=blue](7,-0.5)(7,11.5)
\psline[linewidth=2pt,linecolor=blue](8,-0.5)(8,11.5)
\psline[linewidth=2pt,linecolor=blue](9,-0.5)(9,11.5)
\psline[linewidth=2pt,linecolor=blue](10,-0.5)(10,11.5)
\psline[linewidth=2pt,linecolor=blue](11,-0.5)(11,11.5)
\psline[linewidth=2pt,linecolor=blue](-0.5,-0.5)(11.5,11.5)
\psline[linewidth=2pt,linecolor=blue](0.5,-0.5)(11.5,10.5)
\psline[linewidth=2pt,linecolor=blue](1.5,-0.5)(11.5,9.5)
\psline[linewidth=2pt,linecolor=blue](2.5,-0.5)(11.5,8.5)
\psline[linewidth=2pt,linecolor=blue](3.5,-0.5)(11.5,7.5)
\psline[linewidth=2pt,linecolor=blue](4.5,-0.5)(11.5,6.5)
\psline[linewidth=2pt,linecolor=blue](5.5,-0.5)(11.5,5.5)
\psline[linewidth=2pt,linecolor=blue](6.5,-0.5)(11.5,4.5)
\psline[linewidth=2pt,linecolor=blue](7.5,-0.5)(11.5,3.5)
\psline[linewidth=2pt,linecolor=blue](8.5,-0.5)(11.5,2.5)
\psline[linewidth=2pt,linecolor=blue](9.5,-0.5)(11.5,1.5)
\psline[linewidth=2pt,linecolor=blue](10.5,-0.5)(11.5,0.5)
\psline[linewidth=2pt,linecolor=blue](-0.5,0.5)(10.5,11.5)
\psline[linewidth=2pt,linecolor=blue](-0.5,1.5)(9.5,11.5)
\psline[linewidth=2pt,linecolor=blue](-0.5,2.5)(8.5,11.5)
\psline[linewidth=2pt,linecolor=blue](-0.5,3.5)(7.5,11.5)
\psline[linewidth=2pt,linecolor=blue](-0.5,4.5)(6.5,11.5)
\psline[linewidth=2pt,linecolor=blue](-0.5,5.5)(5.5,11.5)
\psline[linewidth=2pt,linecolor=blue](-0.5,6.5)(4.5,11.5)
\psline[linewidth=2pt,linecolor=blue](-0.5,7.5)(3.5,11.5)
\psline[linewidth=2pt,linecolor=blue](-0.5,8.5)(2.5,11.5)
\psline[linewidth=2pt,linecolor=blue](-0.5,9.5)(1.5,11.5)
\psline[linewidth=2pt,linecolor=blue](-0.5,10.5)(0.5,11.5)
\multirput{0}(0,0)(0,1){12}{%
  \multirput{0}(0,0)(1,0){12}{%
     \pscircle*[linecolor=white]{8pt}
     \pscircle[linewidth=1pt,linecolor=black]{8pt}
   }
}
\rput{0}(0,0){{\bf 1}}
\rput{0}(0,1){{\bf 4}}
\rput{0}(0,2){{\bf 3}}
\rput{0}(0,3){{\bf 4}}
\rput{0}(0,4){{\bf 1}}
\rput{0}(0,5){{\bf 4}}
\rput{0}(0,6){{\bf 2}}
\rput{0}(0,7){{\bf 4}}
\rput{0}(0,8){{\bf 3}}
\rput{0}(0,9){{\bf 4}}
\rput{0}(0,10){{\bf 1}}
\rput{0}(0,11){{\bf 4}}
\rput{0}(1,0){{\bf 2}}
\rput{0}(1,1){{\bf 3}}
\rput{0}(1,2){{\bf 1}}
\rput{0}(1,3){{\bf 2}}
\rput{0}(1,4){{\bf 3}}
\rput{0}(1,5){{\bf 2}}
\rput{0}(1,6){{\bf 1}}
\rput{0}(1,7){{\bf 3}}
\rput{0}(1,8){{\bf 2}}
\rput{0}(1,9){{\bf 1}}
\rput{0}(1,10){{\bf 2}}
\rput{0}(1,11){{\bf 3}}
\rput{0}(2,0){{\bf 1}}
\rput{0}(2,1){{\bf 4}}
\rput{0}(2,2){{\bf 2}}
\rput{0}(2,3){{\bf 4}}
\rput{0}(2,4){{\bf 1}}
\rput{0}(2,5){{\bf 4}}
\rput{0}(2,6){{\bf 3}}
\rput{0}(2,7){{\bf 4}}
\rput{0}(2,8){{\bf 1}}
\rput{0}(2,9){{\bf 4}}
\rput{0}(2,10){{\bf 3}}
\rput{0}(2,11){{\bf 4}}
\rput{0}(3,0){{\bf 2}}
\rput{0}(3,1){{\bf 3}}
\rput{0}(3,2){{\bf 1}}
\rput{0}(3,3){{\bf 3}}
\rput{0}(3,4){{\bf 2}}
\rput{0}(3,5){{\bf 3}}
\rput{0}(3,6){{\bf 2}}
\rput{0}(3,7){{\bf 1}}
\rput{0}(3,8){{\bf 2}}
\rput{0}(3,9){{\bf 3}}
\rput{0}(3,10){{\bf 2}}
\rput{0}(3,11){{\bf 1}}
\rput{0}(4,0){{\bf 3}}
\rput{0}(4,1){{\bf 4}}
\rput{0}(4,2){{\bf 2}}
\rput{0}(4,3){{\bf 4}}
\rput{0}(4,4){{\bf 1}}
\rput{0}(4,5){{\bf 4}}
\rput{0}(4,6){{\bf 1}}
\rput{0}(4,7){{\bf 4}}
\rput{0}(4,8){{\bf 3}}
\rput{0}(4,9){{\bf 4}}
\rput{0}(4,10){{\bf 1}}
\rput{0}(4,11){{\bf 4}}
\rput{0}(5,0){{\bf 1}}
\rput{0}(5,1){{\bf 2}}
\rput{0}(5,2){{\bf 3}}
\rput{0}(5,3){{\bf 1}}
\rput{0}(5,4){{\bf 2}}
\rput{0}(5,5){{\bf 3}}
\rput{0}(5,6){{\bf 2}}
\rput{0}(5,7){{\bf 3}}
\rput{0}(5,8){{\bf 2}}
\rput{0}(5,9){{\bf 1}}
\rput{0}(5,10){{\bf 2}}
\rput{0}(5,11){{\bf 3}}
\rput{0}(6,0){{\bf 2}}
\rput{0}(6,1){{\bf 4}}
\rput{0}(6,2){{\bf 1}}
\rput{0}(6,3){{\bf 4}}
\rput{0}(6,4){{\bf 3}}
\rput{0}(6,5){{\bf 4}}
\rput{0}(6,6){{\bf 1}}
\rput{0}(6,7){{\bf 4}}
\rput{0}(6,8){{\bf 1}}
\rput{0}(6,9){{\bf 4}}
\rput{0}(6,10){{\bf 3}}
\rput{0}(6,11){{\bf 4}}
\rput{0}(7,0){{\bf 3}}
\rput{0}(7,1){{\bf 1}}
\rput{0}(7,2){{\bf 2}}
\rput{0}(7,3){{\bf 3}}
\rput{0}(7,4){{\bf 2}}
\rput{0}(7,5){{\bf 1}}
\rput{0}(7,6){{\bf 2}}
\rput{0}(7,7){{\bf 3}}
\rput{0}(7,8){{\bf 2}}
\rput{0}(7,9){{\bf 3}}
\rput{0}(7,10){{\bf 1}}
\rput{0}(7,11){{\bf 2}}
\rput{0}(8,0){{\bf 1}}
\rput{0}(8,1){{\bf 4}}
\rput{0}(8,2){{\bf 3}}
\rput{0}(8,3){{\bf 4}}
\rput{0}(8,4){{\bf 1}}
\rput{0}(8,5){{\bf 4}}
\rput{0}(8,6){{\bf 3}}
\rput{0}(8,7){{\bf 4}}
\rput{0}(8,8){{\bf 1}}
\rput{0}(8,9){{\bf 4}}
\rput{0}(8,10){{\bf 2}}
\rput{0}(8,11){{\bf 4}}
\rput{0}(9,0){{\bf 2}}
\rput{0}(9,1){{\bf 3}}
\rput{0}(9,2){{\bf 2}}
\rput{0}(9,3){{\bf 1}}
\rput{0}(9,4){{\bf 2}}
\rput{0}(9,5){{\bf 3}}
\rput{0}(9,6){{\bf 2}}
\rput{0}(9,7){{\bf 1}}
\rput{0}(9,8){{\bf 3}}
\rput{0}(9,9){{\bf 2}}
\rput{0}(9,10){{\bf 3}}
\rput{0}(9,11){{\bf 1}}
\rput{0}(10,0){{\bf 3}}
\rput{0}(10,1){{\bf 4}}
\rput{0}(10,2){{\bf 1}}
\rput{0}(10,3){{\bf 4}}
\rput{0}(10,4){{\bf 3}}
\rput{0}(10,5){{\bf 4}}
\rput{0}(10,6){{\bf 1}}
\rput{0}(10,7){{\bf 4}}
\rput{0}(10,8){{\bf 2}}
\rput{0}(10,9){{\bf 4}}
\rput{0}(10,10){{\bf 1}}
\rput{0}(10,11){{\bf 4}}
\rput{0}(11,0){{\bf 2}}
\rput{0}(11,1){{\bf 1}}
\rput{0}(11,2){{\bf 2}}
\rput{0}(11,3){{\bf 3}}
\rput{0}(11,4){{\bf 2}}
\rput{0}(11,5){{\bf 1}}
\rput{0}(11,6){{\bf 3}}
\rput{0}(11,7){{\bf 2}}
\rput{0}(11,8){{\bf 1}}
\rput{0}(11,9){{\bf 3}}
\rput{0}(11,10){{\bf 2}}
\rput{0}(11,11){{\bf 3}}
\multirput{0}(11.5,-0.5)(0,1){12}{%
 \psline[linewidth=2pt,linecolor=black]{->}(0.2,-0.2)(-0.1,0.1)
}
\uput[0](11.7,-0.7){\bf D1}
\uput[0](11.7,0.3){\bf D2}
\uput[0](11.7,1.3){\bf D3}
\uput[0](11.7,2.3){\bf D4}
\uput[0](11.7,3.3){\bf D5}
\uput[0](11.7,4.3){\bf D6}
\uput[0](11.7,5.3){\bf D7}
\uput[0](11.7,6.3){\bf D8}
\uput[0](11.7,7.3){\bf D9}
\uput[0](11.7,8.3){\bf D10}
\uput[0](11.7,9.3){\bf D11}
\uput[0](11.7,10.3){\bf D12}
\endpspicture
\caption{ \label{prop.kag.12k.fig3}
The four-colouring of $T(12,12)$ after Step~4 of the algorithm given
by the proof of Case~2 of Theorem~\protect\ref{theo.kag}
}
\end{figure}
%
%

Finally, every vertex on D$(6k+2)$ has its colour fixed by their neighbours. 
All of them are coloured $2$ except two vertices: the vertex at $x=x_0+2$ is
coloured $3$, and the vertex at $x=x_0+1$ is coloured $1$. 
The sought colouring is depicted in Figure~\ref{prop.kag.12k.fig3}. 
In this case, the increment in the degree is $-2$. Therefore, the degree 
of the final four-colouring is 
\begin{equation}
\deg f \;=\; 6 + 12(k-1) \;\equiv\; 6 \pmod{12} 
\end{equation}
This colouring $f$ of $T(12k,12k)$ is proper, belongs to the restricted set
$\widetilde{\mathcal{C}}_4(T(12k,12k))$ and its degree is congruent 
to six modulo $12$, as claimed. \hfill \qed 

The following Corollary follows trivially:

\begin{corollary} \label{coro.kag}
The single-site dynamics for the three-state Potts antiferromagnet 
at zero temperature on the kagom\'e graph $T'(3L,3L)$ with $L\in\N$ is 
not ergodic.
\end{corollary}

\proof
It follows from the fact that single-site moves are a subset of the Kempe
moves. In fact, each proper 3--colouring of the kagom\'e graph $T'(3L,3L)$
is an ergodicity class in the single-site dynamics. This is a consequence
of each vertex belonging to two neighbouring triangular faces. Therefore, in
every proper 3--colouring, every vertex has two neighboring vertices 
coloured with any of two possible other colours, and no change is possible. 
Therefore, each proper 3--colouring is {\em frozen}\/ in the single-site 
dynamics, and constitutes an ergodicity class. \qed

%
%
\section*{Acknowledgments}

We are indebted to Alan Sokal for his participation on the early stages
of this work, his encouragement and useful suggestions later on, and for
carefully reading the manuscript and suggesting many improvements.
We also wish to thank Eduardo J.S. Villase\~nor for useful discussions,
and Chris L. Henley for correspondence. 

J.S.\ is grateful for the kind hospitality of
the Physics Department of New York University and the Mathematics 
Department of University College London, where part of this work was done.

The authors' research was supported in part by 
the ARRS (Slovenia) Research Program P1--0297, by an NSERC Discovery Grant,
and by the Canada Research Chair program (B.M.), 
by U.S.\ National Science Foundation grant PHY--0424082, 
and by Spanish MEC grants MTM2008--03020 and FPA2009-08785 (J.S.).

%
%


\section*{References}
\begin{thebibliography}{80}

\bibitem{Potts_52}  Potts R B, 1952 {\it Proc. Cambridge Philos. Soc.} 
      {\bf 48} 106.
  
\bibitem{Wu_82} Wu F Y, 1982 \RMP {\bf 54} 235; Erratum 1983 {\bf 55} 315.
  
\bibitem{Wu_84} Wu F Y, 1984 \JAP {\bf 55} 2421.
   
\bibitem{Salas_Sokal_97}  Salas J and Sokal A D, 1997 {\em J. Statist. Phys.}
   {\bf 86} 551  (arXiv:cond-mat/9603068).

\bibitem{Lieb} Aizenman M and Lieb E H, 1981 {\em J. Statist. Phys.}
   {\bf 24} 279.

\bibitem{Chow} Chow Y and Wu F Y, 1987 {\em Phys. Rev. } B {\bf 36} 285.
 
\bibitem{Chalker} Chalker J T, Holdsworth P C W and Shender E F, 1992 
   \PRL {\bf 68} 855. 

\bibitem{Huse} Huse D A and Rutenberg A D, 1992 {\em Phys. Rev.} B 
        {\bf 45} 7536.

\bibitem{Ritchey} Chandra P, Coleman P and Ritchey I, 1993 
        {\em J Physique I} {\bf 3}, 591. 

\bibitem{Henley} Henley C L, 2009 {\em Phys. Rev.} B {\bf 80} 180401 
   (arXiv:0811l.0026).
  
\bibitem{Bremaud} Bremaud P, 1999 {\em Markov Chains, Gibbs Fields, Monte Carlo
  Simulation and Queues}. Texts in Applied Mathematics.
  (Springer-Verlag, Berlin-Heidelberg).

\bibitem{Landau} Landau D P and Binder K, 2009 {\em A Guide to Monte Carlo
  Simulations in Statistical Physics}, 3rd edition. (Cambridge University
  Press, Cambridge). 

\bibitem{WSK_89}  Wang J--S, Swendsen R H and Koteck\'y R, 1989 \PRL 
        {\bf 63} 109. 

\bibitem{WSK_90} Wang J--S, Swendsen R H and Koteck\'y R, 1990 
   {\em Phys. Rev.} B {\bf 42} 2465.

\bibitem{Gibbons} Gibbons A, 1985 {\em Algorithmic graph theory}.
      (Cambridge University Press, Cambridge).

\bibitem{Wagon} Hutchinson J and Wagon S, 1998 {\em Am. Math. Monthly}
       {\bf 105} 170. 

\bibitem{Mohar_05} Mohar B, {\em Kempe equivalence of colorings} 2006
   {\em Graph Theory in Paris, Proc. Conf. in Memory of Claude Berge}\/, 
   ed J A Bondy, J Fonlupt, J L Fouquet, J C Fournier and J 
   J Ram\'{\i}rez Alfons\'{\i}n (Birkhauser, Basel), pp.~287--297. 

\bibitem{Privman} Privman V, {\em Finite--Size Scaling Theory} 1990
       {\em Finite Size Scaling and Numerical Simulations of Statistical 
       Systems}, ed V Privman  (World Scientific, Singapore), pp.~4--98 

\bibitem{Kano_53} Kano K and Naya S, 1953 {\em Prog. Theor. Phys.} {\bf 10} 158.

\bibitem{Mohar_Salas} Mohar B and Salas J, 2009 \JPA {\bf 42} 225204
    (arXiv:0901.1010). 

\bibitem{Fisk_73a} Fisk S, 1973 {\em Adv. Math.} {\bf 11} 326.
   
\bibitem{Fisk_77a} Fisk S, 1977 {\em Adv. Math.} {\bf 24} 298. 
   
\bibitem{Fisk_77b} Fisk S, 1977 {\em Adv. Math.} {\bf 25} 226. 

\bibitem{Kasteleyn_69}  Kasteleyn P W and Fortuin C M, 1969
   {\em J. Phys. Soc. Japan} {\bf 26} (Suppl.) 11.

\bibitem{Fortuin_72}  Fortuin C M and Kasteleyn P W, 1972
   {\em Physica} {\bf 57} 536.

\bibitem{Sokal_bcc2005}  Sokal A D, {\em The multivariate Tutte polynomial
   (alias Potts model) for graphs and matroids} 2005
   {\em Surveys in Combinatorics, 2005}\/, ed B E Webb  
   (Cambridge University Press, Cambridge--New York), pp.~173--226  
   (arXiv:math.CO/0503607).

\bibitem{Altschulter} Altschulter A 1971 {\em Discrete Math.} {\bf 1} 211.

\bibitem{Henley_97a} Burton J K and Henley C L, 1997 \JPA  {\bf 30} 8385
  (arXiv:cond-mat/9708171).

\bibitem{Sokal_99a} Ferreira S J and Sokal A D, 1999 
  {\em J. Stat. Phys.} {\bf 96},461 (arXiv:cond-mat/9811345).

\bibitem{Sokal_93} Lubin M and Sokal A D, 1993 \PRL {\bf 71} 1889. 

\bibitem{Jerrum_private} M. Jerrum, private communication.
\end{thebibliography}
\end{document}